\documentclass[preprint,5p,times,twocolumn]{elsarticle}

\usepackage{amssymb}
\usepackage{amsmath}
\usepackage{multirow}
\usepackage{url}
\usepackage[ruled,vlined]{algorithm2e}
\usepackage{booktabs}   
\usepackage{siunitx}    
\usepackage{tabularx}
\usepackage{adjustbox}
\usepackage{colortbl}
\usepackage[table]{xcolor}
\definecolor{rowgray}{gray}{0.93} 

\usepackage{makecell}
\usepackage{tikz}
\usetikzlibrary{decorations.pathreplacing}

\journal{Medical Image Analysis}

\begin{document}

\begin{frontmatter}



\title{TomoGraphView: 3D Medical Image Classification with \\Omnidirectional Slice Representations and Graph Neural Networks}

\author[label1,label2,label3,label7]{Johannes Kiechle\corref{cor1}}
\author[label1,label2,label3,label7]{Stefan M. Fischer}
\author[label1,label3]{Daniel M. Lang}
\author[label1,label3,label7]{Cosmin I. Bercea}
\author[label6]{\\Matthew J. Nyflot}
\author[label1,label3]{Lina Felsner}
\author[label1,label3,label5,label7]{Julia A. Schnabel\corref{cor2}}
\author[label2,label4]{Jan C. Peeken\corref{cor2}}

\affiliation[label1]{organization={School of Computation, Information and Technology, Technical University of Munich},
            country={Germany}}

\affiliation[label2]{organization={Department of Radiation Oncology, TUM School of Medicine, TUM University Hospital rechts der Isar, Technical University of Munich},
            country={Germany}}

\affiliation[label3]{organization={Institute of Machine Learning in Biomedical Imaging, Helmholtz Munich},
            city={Neuherberg},
            country={Germany}}
            
\affiliation[label4]{organization={Institute of Radiation Medicine, Helmholtz Munich},
            city={Neuherberg},
            country={Germany}}

\affiliation[label5]{organization={School of Biomedical Engineering and Imaging Sciences, King's College London},
            country={United Kingdom}}

\affiliation[label6]{organization={Department of Radiation Oncology, University of Washington},
            city={Seattle},
            country={United States of America}}

\affiliation[label7]{organization={Munich Center for Machine Learning (MCML)},
            city={Munich},
            country={Germany}}


\cortext[cor1]{Corresponding author: johannes.kiechle@tum.de}

\cortext[cor2]{These senior authors contributed equally to this work.}

\begin{abstract}
The sharp rise in medical tomography examinations has created a demand for automated systems that can reliably extract informative features for downstream tasks such as tumor characterization. Although 3D volumes contain richer information than individual slices, effective 3D classification remains difficult: volumetric data encode complex spatial dependencies, and the scarcity of large-scale 3D datasets has constrained progress toward 3D foundation models. As a result, many recent approaches rely on 2D vision foundation models trained on natural images, repurposing them as feature extractors for medical scans with surprisingly strong performance. Despite their practical success, current methods that apply 2D foundation models to 3D scans via slice-based decomposition remain fundamentally limited. Standard slicing along axial, sagittal, and coronal planes often fails to capture the true spatial extent of a structure when its orientation does not align with these canonical views. More critically, most approaches aggregate slice features independently, ignoring the underlying 3D geometry and losing spatial coherence across slices.

To overcome these limitations, we propose \emph{TomoGraphView}, a novel framework that integrates \emph{omnidirectional} volume slicing with \emph{spherical graph-based feature aggregation}. Instead of restricting the model to axial, sagittal, or coronal planes, our method samples both canonical and non-canonical cross-sections generated from uniformly distributed points on a sphere enclosing the volume. Triangulating these viewpoints yields a spherical graph that captures spatial relationships among views, and we use a graph neural network to aggregate their features accordingly. Experiments across six oncology $3$D medical image classification datasets demonstrate that \emph{omnidirectional} volume slicing improves the average performance in Area Under the Receiver Operating Characteristic Curve (AUROC) from $0.7701$ to $0.8154$ compared with traditional slicing approaches relying on canonical view planes. Moreover, we can further improve AUROC performance from $0.8198$ to $0.8372$ by leveraging our proposed graph neural network-based feature aggregation. Notably, \emph{TomoGraphView} surpasses large-scale pretrained $3$D medical imaging models across all datasets and tasks, underscoring its effectiveness as a powerful framework for volumetric analysis and therefore represents a key step toward bridging the gap until fully native $3$D foundation models become available in medical image analysis. We provide a user-friendly library for \emph{omnidirectional} volume slicing at \url{https://pypi.org/project/OmniSlicer}.

\end{abstract}




\end{frontmatter}




\section{Introduction}


Deep learning methods have become a central driver of progress in medical image analysis, owing to their ability to learn complex hierarchical and semantically rich feature representations from large-scale imaging data. These capabilities enable a broad spectrum of clinical applications, ranging from early disease detection and accurate diagnosis~\cite{rana2023machine}, malignancy prediction~\cite{prakash2020breast, naser2020brain, navarro2021development}, and lesion characterization \cite{herent2019detection, yin2023value, dadoun2022deep}, to robust image segmentation~\cite{ronneberger2015u, isensee2021nnu, cao2022swin, erdur2025deep} and prognostic as well as predictive modeling~\cite{zhu2020application, schulz2021multimodal, spohn2023development}. While these capabilities may not only facilitate personalized treatment strategies, they also hold the potential to improve therapeutic effectiveness and patient outcomes~\cite{rasuli2025deep}. As advances in imaging hardware continue to accelerate the acquisition of high-resolution tomographic data, the need for automated systems capable of distilling meaningful imaging biomarkers from complex $3$D volumes to support clinicians has grown correspondingly~\cite{thrall2018artificial}. Effective extraction of such biomarkers is not only essential for alleviating clinician workload but also constitutes a foundational step for enhancing the efficacy of subsequent downstream analytical tasks~\cite{rasuli2025deep}.


One promising direction to extract semantically rich and versatile imaging features from $2$D images lies in the use of vision foundation models, which have demonstrated remarkable generalization capabilities across diverse $2$D computer vision tasks~\cite{radford2021learning, kirillov2023segment, zou2023segment, oquab2023dinov2, fang2024eva, tschannen2025siglip, venkataramanan2025franca, simeoni2025dinov3}. These models excel at producing both substantially meaningful and robust feature embeddings and can be readily adapted to a wide range of downstream tasks, including segmentation~\cite{ayzenberg2024dinov2}, classification~\cite{veasey2024parameter}, or registration~\cite{song2024dino}. However, despite their widespread adoption in $2$D natural image analysis, efforts to translate foundation model principles to $3$D medical imaging have progressed more slowly. A primary constraint remains the absence of large and publicly available multimodal medical imaging datasets required for training at scale~\cite{azad2023foundational}. Building on the criteria expressed by \citeauthor{paschali2025foundation}~\cite{paschali2025foundation}, a radiology foundation model should couple large-scale architectures with multimodal imaging data, rely on self-supervised training to limit annotation burdens, and exhibit emergent capabilities beyond its explicit training objectives. Yet, despite recent progress and rapid development of large-scale pretrained $3$D models, the performance of $3$D radiology foundation models still trails that of state-of-the-art $2$D vision model counterparts~\cite{van2025foundation}. Nevertheless, despite challenges such as the limited availability of large-scale $3$D datasets and the computational demands of training $3$D models, ongoing research in these areas is likely to lead to substantial breakthroughs and enhanced integration of native $3$D models into clinical \mbox{workflows~\cite{azad2023foundational, van2025foundation}}.

Despite the current performance gap between large-scale $2$D and $3$D pretrained models, recent studies have demonstrated that $2$D vision foundation models trained on natural images, particularly the DINO~\cite{caron2021emerging} model family, can be transferred to medical applications with strong results~\cite{huix2024natural, liu2025doesdinov3setnew}. 
Their use, however, necessitates decomposing $3$D medical volumes into $2$D slices, a process that inevitably disrupts the intrinsic spatial coherence of volumetric data.

A common strategy to combine the resulting slice-wise information extracted by a pretrained encoder is to use a dimension-wise average thereof or a shallow multi-layer perceptron (MLP)~\cite{kiechle2024graph, muller2025medical, liu2025doesdinov3setnew}. While this enables end-to-end learning for downstream tasks such as classification, there is no inherent mechanism that explicitly accounts for the relative positions among the slices. More structured approaches attempt to mitigate this limitation: LSTM-based aggregation~\cite{donahue2015long, nguyen2020cnn, lilhore2025hybrid} treats sequential slices analogously to temporal data, capturing inter-slice associations, while transformer-based models~\cite{muller2025medical, meneghetti2025end} leverage self-attention to more directly model spatial relationships across slices. Still, these methods ultimately remain constrained by the inherent loss of geometry introduced during slice-based decomposition. 

With the emergence of graph neural networks (GNNs)~\cite{scarselli2008graph}, effective modeling of naturally graph-structured data has become possible. GNNs are particularly well-suited for problems involving pairwise interactions or modeling complex relational data by encoding relationships in the graph topology~\cite{wu2020comprehensive}. 
In the context of $3$D data, GNNs have demonstrated a strong ability to preserve an underlying spatial structure when the graph topology is derived on the basis of the relative position of $2$D representations~\cite{Wei_2020_CVPR, Wei_2023_IEEE}. In medical imaging, this makes GNNs well-suited to capture spatial relationships between encoded slices, thereby preserving aspects of the underlying $3$D context that are lost in traditional slice-wise representations~\cite{kiechle2024graph, di2025structured}. 

In this work, we introduce \emph{TomoGraphView}, a novel framework that preserves the spatial structure of volumetric medical images while leveraging the representational power of a $2$D foundation model encoder. The name reflects its core components, where \emph{Tomo} refers to tomography data, \emph{Graph} denotes the use of GNN-based feature aggregation, and \emph{View} signifies the integration of multi-view information from several perspectives. More specifically, our framework integrates two key components: \emph{omnidirectional} volume slicing and \emph{spherical graph-based feature aggregation}. First, rather than relying solely on canonical axial, coronal, or sagittal planes, we sample cross-sectional views by means of uniformly distributed points on a sphere surrounding the $3$D volume, capturing a diverse set of canonical and non-canonical perspectives. These viewpoints are triangulated into a spherical mesh that defines the graph topology, with each node corresponding to features extracted from a $2$D vision foundation model applied to a specific slice. Edges between nodes are weighted according to their relative distance, enabling the model to efficiently integrate both local and global context while explicitly preserving spatial structure. By aggregating node features through a GNN, our framework effectively models 3D context from 2D representations, demonstrating the strength of graph-based feature aggregation for volumetric data analysis. The primary contributions of this work are as follows:

\vspace{0.17cm}

\textbf{(I)} We propose a novel out-of-plane slicing technique, termed \emph{omnidirectional} volume slicing, which extends traditional slicing beyond canonical anatomical axes and yields consistently superior downstream performance compared with standard axial–coronal–sagittal approaches.

\vspace{0.17cm}

\textbf{(II)} We introduce a spherical mesh-based graph topology derived from the \emph{omnidirectional} sampling strategy, enabling faithful preservation of spatial relationships between multi-view $2$D cross-sections within a unified \emph{spherical graph-based feature aggregation} framework.

\vspace{0.17cm}

\textbf{(III)} Extensive experiments on six oncology $3$D medical imaging datasets demonstrate that the proposed \emph{TomoGraphView} framework, integrating \emph{omnidirectional} volume slicing with \emph{spherical graph-based feature aggregation}, consistently outperforms slice-wise aggregation baselines and large-scale pretrained $3$D medical models across all datasets and tasks, establishing \emph{TomoGraphView} as a powerful and generalizable framework for volumetric medical image classification.

\section{Related Works}
In the context of preserving the spatial structure of volumetric data when decomposed into $2$D slices compatible with $2$D encoders, we review prior work in two main areas. In Section~\ref{subsec:slice-based}, we begin with $2$D slice-based methods for $3$D medical image analysis, focusing specifically on approaches that extract slices directly from volumetric data. Projection-based methods, such as maximum intensity projections~\cite{antropova2018use, hu2021improved, takahashi2022deep}, which collapse $3$D volumes into $2$D representations, are therefore excluded from this discussion. Subsequently, in Section~\ref{related_work:slice_wise_feature_aggregation}, we turn our attention to slice-wise feature aggregation methods.

\subsection{Slice-based Methods in $3$D Medical Imaging}
\label{subsec:slice-based}

Two-dimensional slice-based approaches remain an appealing alternative to computationally intensive $3$D architectures, as they offer substantially lower memory and compute requirements, benefit from the decomposition of $3$D volumes into numerous $2$D slices, and can leverage powerful pretrained backbones from natural image analysis. These models have repeatedly demonstrated strong performance in medical image classification tasks, underscoring their practical relevance despite the lack of explicit volumetric reasoning~\cite{huix2024natural, liu2025doesdinov3setnew}.

A substantial body of work highlights the effectiveness of $2$D paradigms across diverse clinical settings. \citeauthor{navarro2021development}~\cite{navarro2021development} fine-tuned a DenseNet161 pretrained on ImageNet using axial T1- and T2-weighted MRI slices of soft-tissue sarcomas for tumor grading. Similarly, \citeauthor{ahamed2023slice}~\cite{ahamed2023slice} trained a ResNet18 on axial PET/CT slices of lymphoma to classify whether a slice intersected tumor tissue. \citeauthor{wang2024svfr}~\cite{wang2024svfr} proposed a two-stage slice-to-volume feature representation framework for Alzheimer’s disease, where informative axial MRI slices were selected and aggregated into volume-level features. In parallel, self-supervised foundation models have also been evaluated for slice-level tasks. \citeauthor{huang2024comparative}~\cite{huang2024comparative} and \citeauthor{baharoon2024evaluatinggeneralpurposevision}~\cite{baharoon2024evaluatinggeneralpurposevision} applied DINOv2 to axial brain MRIs and chest CTs for glioma grading and COVID-19 diagnosis, respectively, while \citeauthor{liu2025doesdinov3setnew}~\cite{liu2025doesdinov3setnew} assessed DINOv3 on non-contrast axial CT slices for detecting 18 clinically significant abnormalities.

Building on the limitations of single-plane analysis, several studies have explored multi-plane, or so-called $2.5$D, strategies that integrate axial, coronal, and sagittal views. \citeauthor{saint20222}~\cite{saint20222} employed an Xception model trained on $2.5$D representations to predict HPV status in oropharyngeal cancer. Similarly, \citeauthor{meneghetti2025end}~\cite{meneghetti2025end} explored foundation model-based multiple instance learning for head and neck cancer outcomes, comparing single-plane, multi-plane, and $3$D sub-volume representations. A related hybrid approach by \citeauthor{jang2022m3t}~\cite{jang2022m3t} first extracted volumetric features using a $3$D CNN, followed by multi-plane slicing of the feature tensor and $2$D CNN and transformer processing for Alzheimer’s disease classification.


Despite their empirical success, these methods exclusively rely on canonical anatomical planes aligned with the axial, coronal, and sagittal axes. As a result, structures oriented obliquely or irregularly within the volume may be inadequately represented or omitted entirely. Moreover, beyond the choice of slicing strategy, the mechanism used to aggregate slice-wise features is equally crucial for preserving volumetric context. How these features are combined ultimately determines the extent to which spatial coherence is retained, motivating a deeper examination of aggregation strategies in the subsequent section.

\begin{figure*}[t]
\centering
\includegraphics[width=1.0\linewidth]{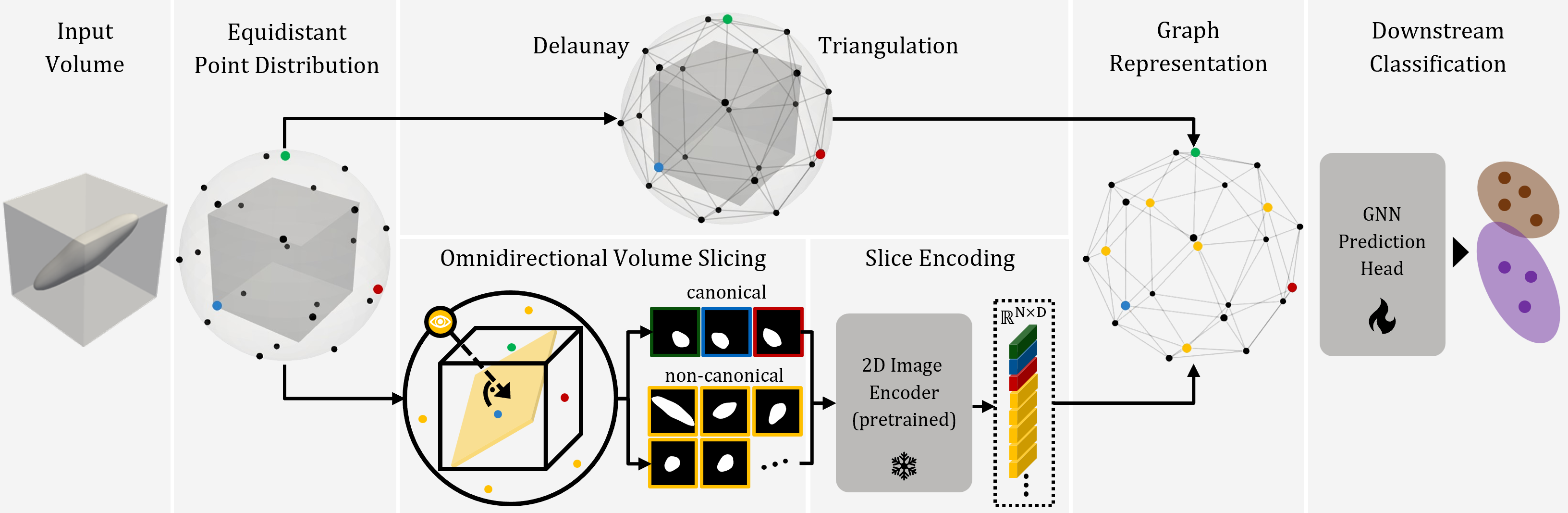}
\caption{Visual overview of the proposed \emph{TomoGraphView} framework. The input volume is enclosed within a sphere with $N$ equidistant surface points, three corresponding to canonical planes indicated as green (axial), blue (coronal), red (sagittal), and the remaining $N-3$ distributed via repulsion-based optimization. From this distribution, two processing branches emerge: Delaunay triangulation forms a spherical mesh defining the graph topology (done only once, for any $N$ node graph topology), and \emph{omnidirectional} volume slicing generates canonical and non-canonical slices along the directions of the sampled points. The resulting slices are encoded into $1$D vector representations that serve as node features within the spherical graph. A GNN prediction head performs the downstream classification.}
\label{fig:method}
\end{figure*}

\subsection{Slice-wise Feature Aggregation in $3$D Medical Imaging} \label{related_work:slice_wise_feature_aggregation}

Slice-wise feature aggregation has emerged as a promising paradigm for $3$D medical image analysis, offering a means to leverage powerful $2$D foundation models while circumventing the computational burden of end-to-end $3$D training. However, because slicing inherently discards volumetric structure, downstream performance depends on how effectively inter-slice spatial dependencies are modeled.


The most straightforward strategy involves collapsing slice embeddings into a single volume-level descriptor using non-learnable operations, such as mean pooling. For example, \citeauthor{muller2025medical}~\cite{muller2025medical} encoded axial slices with DINOv2 and averaged the embeddings for breast MRI, lung CT, and knee MRI classification, while \citeauthor{liu2025doesdinov3setnew}~\cite{liu2025doesdinov3setnew} similarly averaged DINOv3 features from axial chest CT slices for anatomical classification. Although computationally efficient, these approaches ignore inter-slice relationships and therefore provide limited volumetric context.

Learnable fusion mechanisms offer a natural extension. Multi-layer perceptrons (MLPs) have been widely adopted to integrate slice-level features, as in \citeauthor{truong2021transferable}~\cite{truong2021transferable}, who fine-tuned self-supervised ImageNet models with an MLP head for several medical classification tasks. \citeauthor{baharoon2024evaluatinggeneralpurposevision}~\cite{baharoon2024evaluatinggeneralpurposevision} scaled this evaluation to 200 radiology experiments across diverse modalities (X-ray, CT, and MRI), and \citeauthor{kiechle2024graph}~\cite{kiechle2024graph} benchmarked MLP aggregation across MedMNIST3D structures. Despite their flexibility, MLPs typically lack explicit awareness of slice order or spatial layout unless augmented with positional conditioning.

Recurrent architectures address this limitation by treating slices as ordered sequences. LSTM-based aggregation has been shown to enhance volumetric reasoning, as demonstrated by \citeauthor{zhang2019rsanet}~\cite{zhang2019rsanet} for multiple sclerosis lesion segmentation and by \citeauthor{nguyen2020cnn}~\cite{nguyen2020cnn} for intracranial hemorrhage detection. Bidirectional variants have also been explored, such as the EfficientNet-B0 + BiLSTM hybrid for breast cancer classification~\cite{lilhore2025hybrid}. Slice-by-slice CNN–LSTM hybrids further illustrate the utility of sequential modeling for capturing long-range spatial dependencies~\cite{miron2021evaluating}.


Transformer-based models provide an alternative that leverages attention to learn global inter-slice relationships. The M3T framework~\cite{jang2022m3t} fused multi-plane representations for Alzheimer’s diagnosis, while \citeauthor{meneghetti2025end}~\cite{meneghetti2025end} introduced a foundation model–based multiple instance learning (MIL) transformer for predicting clinical outcomes in head and neck cancer, encoding $2$D CT slices with a foundation model and aggregating them using transformer-based attention. More recently, the Medical Slice Transformer (MST)~\cite{muller2025medical} combined transformer aggregation with self-supervised $2$D encoders, achieving strong results across breast MRI, lung CT, and knee MRI classification downstream tasks.

A complementary direction represents volumetric data as structured graphs. \citeauthor{kiechle2024graph}~\cite{kiechle2024graph} and \citeauthor{di2025structured}~\cite{di2025structured, dipiazza2025structuredspectralgraphrepresentation} modeled tomographic volumes where nodes correspond to slice features and edges encode spatial adjacency. \citeauthor{kiechle2024graph}~\cite{kiechle2024graph} showed that graph representations improve accuracy and runtime over MLP-based aggregation but depend heavily on selecting suitable graph convolution operators and topologies, limiting general applicability. \citeauthor{di2025structured}~\cite{di2025structured, dipiazza2025structuredspectralgraphrepresentation} constructed graphs from axial slice stacks to capture both local and global inter-slice dependencies along the axial axis. However, this design remains restricted to a single orientation, omitting valuable coronal and sagittal context. Moreover, evaluation limited to chest CTs leaves the generalizability of this approach across modalities and anatomical sites uncertain.

\section{Methodology}

In this work, we present \emph{TomoGraphView}, a novel framework that aims to preserve the spatial structure of volumetric medical imaging data while leveraging the representational power of a $2$D foundation model encoder. 
Generally, our framework is built upon two key components: \emph{omnidirectional} volume slicing and \emph{spherical mesh–based graph topology} construction with corresponding GNN-based feature aggregation and downstream prediction. A visual summary of the overall framework workflow is depicted in Figure \ref{fig:method}. 

The following sections detail each component of our \emph{TomoGraphView} framework. We first describe \emph{omnidirectional} volume slicing in Section \ref{sec:volume_slicing} and relate it to conventional slicing approaches. Section \ref{sec:image_encoding} outlines the encoding of the obtained \emph{omnidirectional} $2$D cross-sectional slices into $1$D feature representations. In Section \ref{sec:graph_toplogy_construction}, we introduce the \emph{spherical mesh–based graph topology} construction, followed by Section \ref{sec:graph_rep_learning}, which describes the graph representation learning process used for downstream classification.

\subsection{Omnidirectional Volume Slicing} \label{sec:volume_slicing}

\begin{figure*}[t]
\centering
\includegraphics[width=1.0\linewidth]{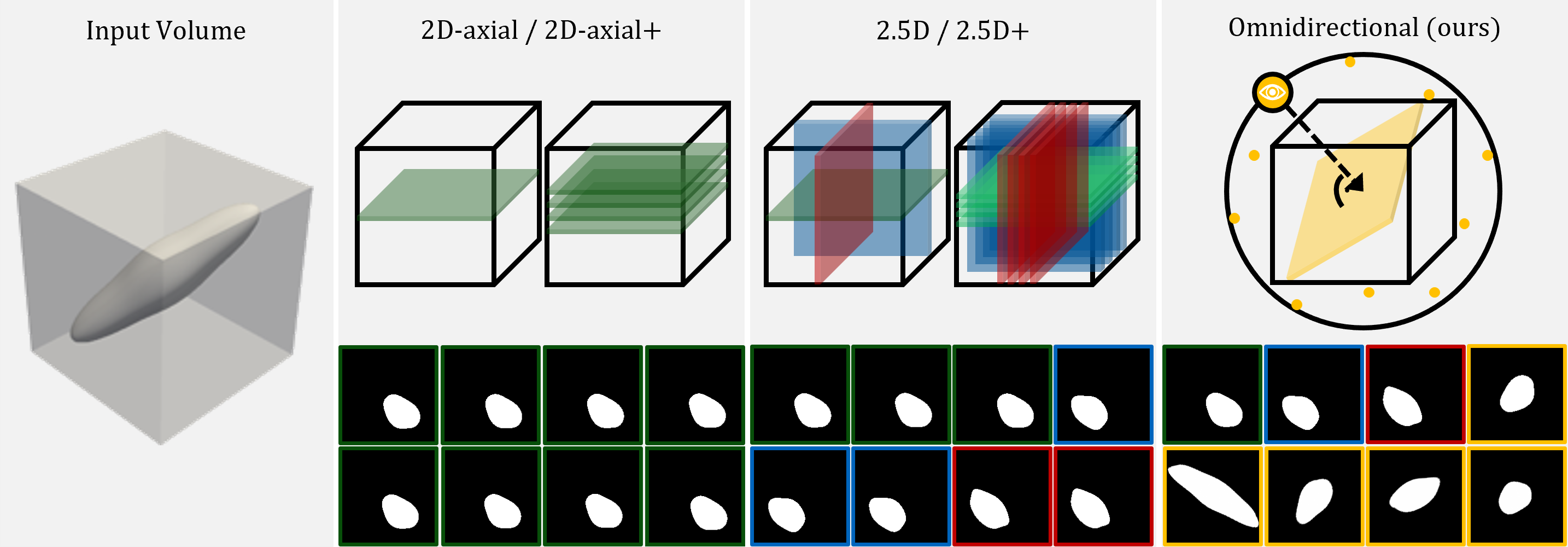}
\caption{Visual representation of different volume slicing strategies and resulting slices for a synthetically generated input volume depicting an elongated shape diagonally within the volume (i.e., not aligned with any canonical axis). The presented volume slicing strategies include $2$D-axial and $2$D-axial+ (left), $2.5$D and $2.5$D+ (middle), and our proposed \emph{omnidirectional} volume slicing (right). Green slices represent slices extracted from the axial direction, red from the sagittal direction, and blue from the coronal direction, whereas yellow shows out-of-plane slices as obtained by means of our \emph{omnidirectional} volume slicing strategy.}
\label{fig:volume_slicing}
\end{figure*}

Traditional volume slicing strategies typically rely on canonical planes such as axial, coronal, and sagittal views, or a combination thereof, referred to as $2.5$D. However, before decomposing a volume into $2$D slices, it is common practice in $3$D medical image classification to first extract a subvolume enclosing the target structure of interest, commonly defined using a segmentation mask. Accordingly, the most representative slice along each direction is defined as the one containing the largest visible lesion area. As illustrated in Figure~\ref{fig:volume_slicing}, in the $2$D-axial setting, a single slice is extracted along the axial direction. The $2$D-axial+ variant further includes adjacent slices symmetrically around the axial largest-lesion slice, resulting in a total of $N$ views. The $2.5$D approach combines information from all the canonical planes by selecting the most representative slice along the axial, coronal, and sagittal directions, whereas $2.5D$+ extends this strategy by additionally including adjacent slices symmetrically around the largest-lesion slice in each direction. 

\begin{figure*}[t]
\centering
\includegraphics[width=1.0\linewidth]{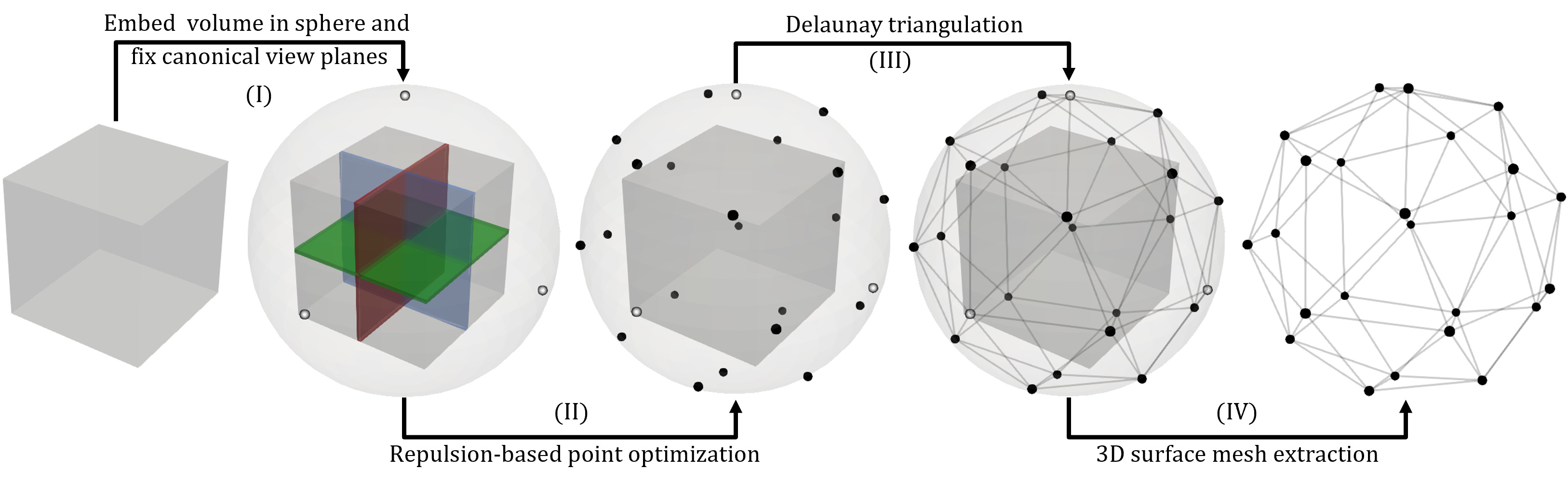}
\caption{Sequential overview of our spherical mesh-based graph topology construction: (I) We fix canonical view planes with coordinates $(1,0,0), (0,1,0)$ and $(0,0,1)$ on the surface of the sphere, (II) we add additional $(N-3)$ points on the sphere and optimize their position for equidistant distribution across the surface, (III) we apply Delaunay triangulation to establish edge connection between neighboring points, and (IV) extract the $3$D surface mesh. }
\label{fig:graph_construction}
\end{figure*}

In this work, additionally to utilizing information from canonical planes, we propose to extract slices from non-canonical orientations to obtain \emph{omnidirectional} slice representations, thereby enabling a richer characterization of volumetric structures that are not naturally aligned with the canonical axes. To obtain $N$ \emph{omnidirectional} slice representations, we visually embed the volume within a bounding sphere $\mathcal{S}$, centered at the origin, which is formally described as
\begin{equation}
    \mathcal{S} = \{ \mathbf{x} \in \mathbb{R}^3 : \|\mathbf{x}\|_2 \leq r \},
\end{equation}
where $r$ is the radius such that the sphere fully contains the $3$D volumetric scan as illustrated in Figure~\ref{fig:graph_construction}. Next, as we intend to make use of both, canonical and non-canonical perspectives, we first of all fix three points on the surface of the sphere representing the canonical view planes (i.e., axial, coronal, and sagittal) with their known coordinates $(r, 0, 0), (0,r,0)$ and $(0,0,r)$, as illustrated in Figure~\ref{fig:graph_construction} (step I).
We then optimize the location of the remaining $(N-3)$ points, representing non-canonical view planes, such that all points on the surface are approximately equally distributed, as illustrated in Figure~\ref{fig:graph_construction} (step II). To this end, we make use of the repulsion-based optimization outlined in Algorithm~\ref{algorithm:repulsion-based_optimization}, which is inspired by the Thomson problem \cite{thomson1904xxiv}. Therein, points are interpreted as identical charged particles constrained to lie on the spherical surface, and their positions are optimized by minimizing the system's total electrostatic energy. Formally, given $N$ points $\{x_i\}_{i=1}^N$ with $x_i \in \mathbb{R}^3$ and $\|x_i\| = 1$, the pairwise simplified Coulomb energy, assuming unit charges and ignoring the Coulomb constant, is used without loss of generality, and defined as
\begin{equation}
E (N) = \sum_{i \neq j} \frac{1}{\|x_i - x_j\|^2}.
\end{equation} \label{eq:pairwise_coulomb}
The gradient of the Coulomb energy function $\nabla E(N)$ yields repulsive forces between all point pairs which is used to iteratively update the position of the $N-3$ non-fixed points using gradient descent. After each update step, points are re-projected onto the sphere to enforce the unit-norm constraint.
%
%
By visually enclosing the volume in a sphere, both centered at the origin, we guarantee that all possible slice orientations, which are defined by points on the sphere, will intersect the volume. For each volume, canonical and non-canonical slices are then extracted by first aligning the slicing plane such that its normal vector points to the respective point sampled on the surface of the sphere, and second by identifying the largest lesion slice from the respective canonical and non-canonical plane. Repeating this process for all $i=1,\dots, N$ yields a set of $N$ uniformly distributed \emph{omnidirectional} slices, thereby effectively resembling the spatial structure of the input volume.

\begin{algorithm}[t]
\caption{Repulsion-based point optimization}
\vspace{0.8em} 
\KwIn{$N$ total points, $F$ fixed points on sphere}
Place fixed points at given coordinates\;
Randomly sample $(N - F)$ additional points on sphere\;
\vspace{0.8em} 
\While{not converged}{
    \ForEach{point $i$}{
        \If{$i$ is fixed}{
            continue\;
        }
        Compute Coulomb force between pairs $j \neq i$\;
        \hskip1.5em $E_i \gets \sum_{j \neq i} \frac{1}{\|x_i - x_j\|^2}$\;
        Update position: $x_i \gets x_i - \eta \cdot \nabla E_i$\;
        Reproject: $x_i \gets x_i / \|x_i\|$\;
    }
}
\vspace{0.8em} 
\KwOut{Optimized set of $N$ points on the sphere}
\vspace{0.8em} 
\label{algorithm:repulsion-based_optimization}
\end{algorithm}

\subsection{Cross-sectional Image Encoding} \label{sec:image_encoding}
After \emph{omnidirectional} volume slicing, we transfer the obtained $2$D volume slices into $1$D feature representations by means of \mbox{DINOv2}~\cite{oquab2023dinov2}. Generally, the DINO~\cite{caron2021emerging} framework is a self-supervised learning method based on a student–teacher architecture designed to extract semantically rich and meaningful visual representations. Building on this, \citeauthor{oquab2023dinov2}~\cite{oquab2023dinov2} introduced DINOv2, an extension of DINO \cite{caron2021emerging} trained with self-supervised contrastive learning on a large-scale dataset of 142 million curated images. 
In this work, we employ a pretrained DINOv2 vision transformer (ViT) model in its small variant, featuring $21$M parameters. The encoder is kept frozen during all experiments and processes input slices of size $224 \times 224$ pixels, obtained by linearly upscaling the input images if they are not initially matched. The model produces a $384$-dimensional feature representation for a single input image corresponding to a single view.

\subsection{Spherical Mesh-based Graph Topology Construction} \label{sec:graph_toplogy_construction}

Given the $N$ view points, which are equidistantly distributed on the spherical surface (Section~\ref{sec:volume_slicing}) and their corresponding cross-sectional volume slice encodings (Section~\ref{sec:image_encoding}), we proceed by describing the construction of the \emph{spherical mesh-based graph topology}. Therein, nodes represent canonical and non-canonical view planes and slice encodings their corresponding feature vector. We denote the set of points uniformly distributed on the spherical surface as $\mathbf{P} = \{\mathbf{p}_1, \dots, \mathbf{p}_N\}$. To derive a graph structure from the sampled points, we apply Delaunay triangulation to $\mathbf{P}$, which guarantees that no point lies inside the circumcircle of any spherical triangle \cite{delaunay1934sphere}. The triangulation defines the edge set
%
\begin{equation}
    \mathbf{E} = \{\, (i, j) \mid \mathbf{p}_i, \mathbf{p}_j \text{ are adjacent in the triangulation} \,\}\,,
\end{equation}
yielding an undirected graph $\mathbf{G} = (\mathbf{V}, \mathbf{E}, \mathbf{A})$, 
where the vertices $\mathbf{V} = \mathbf{P}$ and $\mathbf{A} \in \mathbb{R}^{N \times N}$ is the weighted adjacency matrix, where all connections $\mathbf{e}_{ij}$ resulting from the triangulation have an edge weighting $\mathbf{A}_{ij} = \mathbf{w}_{ij} = 1$. This procedure results in a mesh-based graph topology with near-uniform node distribution and well-defined local neighborhoods. Two properties that have been proven to be advantageous for subsequent mesh-based graph learning tasks \cite{bronstein2017geometric}. A graphical visualization of the Delaunay triangulation (step III) and $3$D surface mesh extraction (step IV) can be found in Figure~\ref{fig:graph_construction}. Besides the local connectivity pattern, where points $\mathbf{p}_i$ and $\mathbf{p}_j$ are connected when they are adjacent in the triangulation, we also introduce weighted cross-connections between all remaining node pairs, effectively forming a complete graph. The corresponding edge weights $\mathbf{w}_{ij}$ of cross-connections are defined as the inverse node distance. Here, the distance refers to the hop distance (i.e., shortest path length between two nodes $(\mathbf{v}_i, \mathbf{v}_j \in \mathbf{V}$) using Dijkstra algorithm \cite{dijkstra2022note}, defined as
\begin{equation}
    \mathbf{d_G}(\mathbf{v}_i, \mathbf{v}_j) = \min_{\mathbf{p} \in \mathbf{P}(\mathbf{v}_i, \mathbf{v}_j)} |\mathbf{p}|\,,
\end{equation}
where $\mathbf{p} \in \mathbf{P}(\mathbf{v}_i, \mathbf{v}_j)$ is the set of all paths from node $\mathbf{v}_i$ to node $\mathbf{v}_j$, and $|\mathbf{p}|$ refers to the number of edges in path $\mathbf{p}$, which shows the minimum distance from node $\mathbf{v}_i$ to node $\mathbf{v}_j$. Thus, $\mathbf{d_G}(\mathbf{v}_i, \mathbf{v}_j)$ corresponds to the minimum number of hops required to reach node $\mathbf{v}_j$ from node $\mathbf{v}_i$. This finally results in edge weights computed as

\begin{equation}
\mathbf{w}_{ij} = \frac{1}{\mathbf{d_G}(\mathbf{v}_i, \mathbf{v}_j)} = \frac{1}{\left (\min_{\mathbf{p} \in \mathbf{P}(\mathbf{v}_i, \mathbf{v}_j)} |\mathbf{p}|\right)}
\end{equation} \label{eq:edge_weights_hop_distance}


\subsection{Graph Representation Learning} \label{sec:graph_rep_learning}

Let $\mathbf{G} = (\mathbf{V}, \mathbf{E}, \mathbf{A})$ denote a spherical mesh-based graph, as introduced in Section~\ref{sec:graph_toplogy_construction}, with node attributes $\mathbf{X_v}$ for $\mathbf{v} \in \mathbf{V}$ and edge weights $\mathbf{w}_{i,j} = \mathbf{A}_{ij}$ for $(i, j) \in \mathbf{E}$. Given a set of graphs $\{\mathbf{G}_N\}$ and their labels $\{\mathbf{y_N}\}$, the task of graph supervised learning is to learn a representation vector $\mathbf{h_G} \in \mathbb{R}^n$ that is used to predict the class of the entire graph. Generally, graph representation learning can be derived from three different flavors of GNN layers, namely convolutional, attentional, and message-passing~\cite{bronstein2021geometric}. In this work, we focus on GNNs that follow message-passing, as it provides a flexible and conceptually intuitive framework where node features are aggregated among neighborhoods, following the underlying graph structure~\cite{gilmer2017neural}. For a single iteration, this is generically represented as follows 
\begin{equation}
    \mathbf{h}_i = \phi \left( \mathbf{x}_i, \bigoplus_{j \in \mathcal{N}_i} \psi(\mathbf{x}_i, \mathbf{x}_j) \right)\,,
\end{equation}
%
where $\mathbf{x}_i$ denotes the feature vector of node $\mathbf{v}_i$, and $\mathcal{N}_i$ represents its set of neighboring nodes. The function $\psi(\mathbf{x}_i, \mathbf{x}_j)$ computes a message from neighbor $j$ to node $i$, and the operator $\bigoplus$ aggregates all incoming messages using a permutation-invariant function. The update function $\phi$ combines the $i$-th node feature $\mathbf{x}_i$ with the aggregated neighborhood representation to produce the updated embedding $\mathbf{h}_i$. More specific for GraphSAGE~\cite{hamilton2017inductive}, the updated node representation $\mathbf{h}_i$ can be described as
\begin{equation} \label{eq:sage}
    \mathbf{h}_i = \phi \left( \mathbf{W} \left[ \mathbf{x}_i \: \|  \:\bigoplus_{j \in \mathcal{N}_i} \left( \left\{\mathbf{e}_{ji} \cdot \mathbf{x}_j, \: \forall j \in \mathcal{N}(i) \right\} \right) \right] \right)\enspace,
\end{equation}
%
where $\mathbf{W}$ is a learnable weight matrix, $\|$ represents the concatenation operation, and 
$\mathbf{e}_{ji}$ is the scalar weight on the edge from node $j$ to node $i$. As using multiple node neighborhood aggregation functions has been shown to provide notable improvements \cite{corso2020principal}, we leverage both mean and max aggregation. They are particularly suitable to capture the distribution of elements within the graph (i.e., mean) and prove to be advantageous to identify representative elements (i.e, max) \cite{xu2018powerful}. As our underlying graph topology does not change within a setting of $N$ nodes, we do not use sum aggregation, which has been shown to be especially powerful for learning structural graph properties. After the node updates, the resulting set of $N$ one-dimensional node feature vectors is aggregated using dimension-wise mean pooling $\delta$ (i.e., a mean readout function) to obtain a global graph representation $\mathbf{h}_G \in \mathbb{R}^{N}$. This representation is then passed through a linear layer $\sigma$ to predict the binary class label $\mathbf{y}_G$ for the entire graph, formally expressed as

\begin{equation}
    \mathbf{y}_G = \sigma(\mathbf{h}_G), \quad 
    \text{where} \quad 
    \mathbf{h}_G = \delta \!\left( \{\, \mathbf{h}_v \mid v \in G \,\} \right).
\end{equation}
        


\begin{figure*}[t]
\centering
\includegraphics[width=1.0\linewidth]{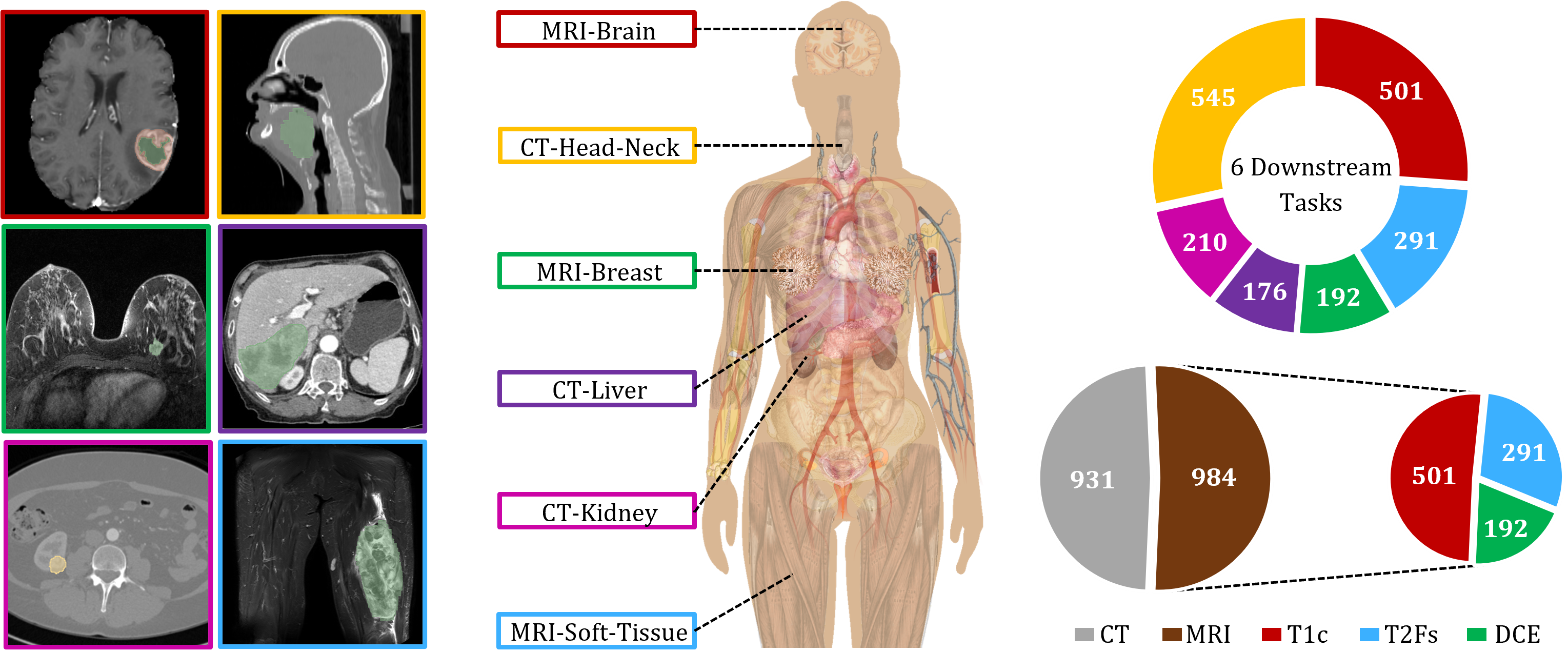}
\caption{Visual summary of the collected datasets from various target anatomical structures across the human body. The overview includes a visual example from each dataset (left), as well as the number of images under different acquisition criteria (right). Here, CT refers to computed tomography images, MRI refers to magnetic resonance imaging, T1c refers to T1-weighted post-contrast, T2Fs refers to T2-weighted fat-saturated, and DCE refers to dynamic contrast-enhanced.}
\label{fig:datasets}
\end{figure*}

\section{Experiments \& Results}

Our experimental evaluation begins with a description of the datasets used in this study (Section \ref{sec:datasets}), followed by details on implementation and training procedures (Section \ref{sec:implementation_and_training_details}). We then assess our proposed volume-slicing strategy (Section \ref{subsec:volume_slicing_benchmarking}), before evaluating the full \emph{TomoGraphView} framework (Section \ref{sec:tomographview_evaluation}). The analysis is extended by examining the effect of slice spacing (Section \ref{subsec:impact_volume_slice_spacing}), conducting a graph-topology ablation (Section \ref{sec:graph_topology_ablation}), and comparing alternative slice-wise feature aggregation methods (Section \ref{sec:slice-wise_feature_aggregation_methods}). Finally, we benchmark \emph{TomoGraphView} against $3$D pretrained models (Section \ref{sec:comparison_3D_models}), considering both frozen and finetuned $3$D backbone configurations.

\subsection{Datasets} \label{sec:datasets}

We focus our experimental evaluation on six clinically meaningful oncological tasks. An overview of the datasets used is summarized below, and Figure 4 provides a visual depiction.

\vspace{0.6em}  

\noindent \textbf{Brain Tumors~\cite{calabrese2022university}} This open-source dataset comprises $n = 501$ patients with brain gliomas, with a median voxel spacing of $1.00 \times 1.00 \times 1.00$ mm. It contains T1-weighted post-contrast (T1c) MRI scans from patients diagnosed with glioblastoma, oligodendroglioma, or astrocytoma. The cohort was collected at a single institution, the University of California, San Francisco, USA. For our analysis, tumors were categorized into high-grade (G4) and low-grade (G2/G3) according to the WHO CNS classification.

\vspace{0.6em}  

\noindent \textbf{Head-Neck Tumors~\cite{wee2019data, grossberg2018imaging, kwan2019data}} This open-source dataset comprises CT scans of $n = 545$ patients with head and neck squamous cell carcinoma (HNSCC) of the oropharynx, acquired at a median voxel spacing of $0.98 \times 0.98 \times 2.00$ mm. The multi-institutional cohort was assembled from Maastricht University Medical Center, Netherlands, University of Texas MD Anderson Cancer Center, USA, and Princess Margaret Cancer Center, Ontario, Canada. The task involves binary classification of human papillomavirus (HPV) status into HPV-positive and HPV-negative entities.

\vspace{0.6em}  

\noindent \textbf{Breast Tumors~\cite{saha2021dynamic}} This open-source dataset includes dynamic contrast-enhanced (DCE) MRI scans of $n = 192$ patients with biopsy-confirmed invasive breast cancer, acquired at a median voxel spacing of $0.74 \times 0.74 \times 1.00$ mm. The single-institutional cohort was retrospectively collected at Duke Hospital, North Carolina, USA. Tumors were stratified into high-grade (“high”) and low-grade (“intermediate”/“low”) categories according to their Nottingham grade. In our experiments, we use the first post-contrast DCE-MRI volumes.

\vspace{0.6em}  

\noindent \textbf{Liver Tumors~\cite{luo2025comprehensive}} This open-source dataset consists of contrast-enhanced multi-phase CT scans from $n = 176$ patients with primary liver tumors, acquired at a median voxel spacing of $1.00 \times 1.00 \times 1.00$ mm. The cohort was collected at the Radiology Department of Chongqing Yubei District People’s Hospital, China. Binary classification was performed on the venous phase (C2) to distinguish hepatocellular carcinoma (HCC) from combined hepatocellular–cholangiocarcinoma (cHCC-CCA), the latter being clinically more aggressive due to poorer overall survival at comparable stage. 

\vspace{0.6em}  

\noindent \textbf{Kidney Tumors~\cite{heller2019kits19}} This open-source dataset contains contrast-enhanced multi-phase CT scans of $n = 210$ patients with kidney tumors, acquired at a median voxel spacing of $0.78 \times 0.78 \times 3.0$ mm. The single-institutional cohort was retrospectively collected at the University of Minnesota, USA. Tumors were graded into high-grade (G3/G4) and low-grade (G1/G2) entities according to the WHO ISUP classification, as determined from post-operative surgical pathology reports.

\vspace{0.6em}  

\noindent \textbf{Soft-Tissue Tumors} This in-house dataset comprises MRI scans of $n = 291$ patients with soft-tissue sarcomas (STS) of the extremities and trunk, acquired with T2-weighted fat-saturation (T2FS) sequences at a median voxel spacing of $0.78 \times 0.94 \times 4.80$ mm. Two independent patient cohorts were retrospectively collected at the University of Washington/Seattle Cancer Care Alliance, USA, and the Technical University of Munich, Germany. Tumors were categorized into high-grade (G2/G3) and low-grade (G1) entities based on histopathological grading.

\subsection{Implementation and Training Details} \label{sec:implementation_and_training_details}

All volumes were first aligned to a common positive Right–Anterior–Superior (RAS+) orientation and subsequently resampled to an isotropic voxel spacing of $1 \times 1 \times 1$ mm. In order to minimize distortion when resizing slices to match the DINOv2 input resolution, we applied symmetric cropping around the lesion. For our proposed \emph{omnidirectional} slicing strategy, scalar image volumes were resampled using linear interpolation, while segmentation masks were resampled with nearest-neighbor interpolation to preserve their categorical labels. For all experiments, we attach a lightweight classification head to the frozen DINOv2 feature extractor. Specifically, the head of each baseline method consists of two layers with a non-linear ReLU activation in between, featuring a total of $100$k trainable parameters to ensure a fair comparison across methods. We restrict the prediction head to a shallow design to avoid overfitting, given the limited dataset sizes in medical imaging, which also helps to mitigate graph oversmoothing in the case of GNNs \cite{keriven2022not}. For the MLP, slice encodings are concatenated. We train and evaluate our models using a 5-fold cross-validation scheme, and stratify splits according to the binary class label. More specifically, we use three splits for training, one for validation, and one for testing. Training is conducted using the SGD optimizer with a weight decay of $1e^{-3}$, a batch size of $16$, and $300$ epochs. The model that achieves the best AUROC score on the validation set is selected for final evaluation on the test set. A learning rate scheduler first linearly increases the learning rate to $0.001$ for $100$ warm-up epochs, and decays the learning rate afterwards by a factor of $0.95$ when there is no improvement in AUROC validation score for at least 5 epochs, to ensure smooth convergence. All models were trained on an Nvidia RTX A6000 GPU. We publicly share our code repository\footnote{\url{http://github.com/compai-lab/2025-MedIA-kiechle}} and furthermore provide a user-friendly library for \emph{omnidirectional} volume slicing\footnote{\url{https://pypi.org/project/OmniSlicer}}.

\begin{table*}[t]
\centering
\caption{Quantitative comparison of conventional volume slicing strategies (2D-axial, 2D-axial+, 2.5D, 2.5D+) with our proposed omnidirectional slicing approach using an MLP prediction head, as well as our TomoGraphView framework. All methods are evaluated across multiple datasets and numbers of volume slices (views) using 5-fold cross-validation and Area Under the Receiver Operating Characteristic Curve (AUROC) as the performance metric. The rightmost column reports the mean performance over all datasets. Bold values indicate the best result per dataset, and underlined values denote the second-best.}

\begin{tabularx}{\textwidth}{X|c|cccccc|c}
\toprule
\textbf{Method (Head -- Slicing Strategy)} 
& \textbf{Views} 
& \textbf{Breast} & \textbf{Brain} & \textbf{Head-Neck} 
& \textbf{Kidney} & \textbf{Liver} & \textbf{Soft-Tissue} & \textbf{Mean} \\
\midrule

\multirow{1}{*}{MLP -- 2D-axial Slicing} 
& 1  & 0.5844 & 0.9013 & 0.6192 & 0.6541 & 0.8821 & 0.7421 & 0.7305 \\
\midrule

\multirow{4}{*}{MLP -- 2D-axial+ Slicing}
& 3  & 0.6166 & 0.8936 & 0.6550 & 0.6234 & 0.8803 & 0.7322 & 0.7335 \\
& 8  & 0.6402 & 0.9164 & 0.6171 & 0.7689 & 0.8738 & 0.7530 & 0.7615 \\
& 16 & 0.6467 & 0.9104 & 0.5982 & 0.7348 & 0.9053 & 0.7555 & 0.7584 \\
& 24 & 0.6299 & 0.9295 & 0.6277 & 0.7806 & 0.8895 & 0.7634 & 0.7701 \\
\midrule

\multirow{1}{*}{MLP -- 2.5D Slicing}
& 3  & 0.6783 & 0.9062 & 0.6818 & 0.6815 & \underline{0.9662} & 0.6951 & 0.7681 \\
\midrule

\multirow{3}{*}{MLP -- 2.5D+ Slicing}
& 8  & 0.7002 & 0.9296 & 0.6555 & 0.7300 & 0.9484 & 0.7488 & 0.7854 \\
& 16 & \textbf{0.7437} & 0.9226 & 0.6588 & 0.7606 & \textbf{0.9667} & 0.7473 & 0.7999 \\
& 24 & 0.6991 & 0.9265 & 0.6799 & 0.7041 & 0.9468 & 0.7751 & 0.7885 \\
\midrule

\multirow{3}{*}{MLP -- Omnidirectional Slicing}
& 8  & 0.6627 & 0.9457 & 0.6790 & 0.7306 & 0.9372 & 0.7702 & 0.7876 \\
& 16 & 0.6689 & 0.9491 & 0.6999 & 0.7450 & 0.9352 & 0.8090 & 0.8012 \\
& 24 & 0.6794 & 0.9492 & \underline{0.7384} & 0.7229 & 0.9421 & \textbf{0.8602} & 0.8154 \\
\midrule \midrule

\rowcolor{rowgray}
& 8  & \underline{0.7336} & \textbf{0.9678} & 0.6780 & \underline{0.7878} & 0.9600 & 0.8399 & \underline{0.8279} \\

\rowcolor{rowgray}
& 16 & 0.6907 & \underline{0.9666} & 0.7244 & 0.7479 & 0.9585 & 0.8389 & 0.8212 \\

\rowcolor{rowgray}
\multirow{-3}{*}{\makecell{GNN -- Omnidirectional Slicing\\\textbf{TomoGraphView (ours)}}}
& 24 & 0.7084 & 0.9655 & \textbf{0.7527} & \textbf{0.7893} & 0.9634 & \underline{0.8441} & \textbf{0.8372} \\

\bottomrule
\end{tabularx}

\label{tab1}
\end{table*}

\subsection{Impact of Volume Slicing Strategy} \label{subsec:volume_slicing_benchmarking}

We start our experimental evaluation by comparing baseline volumetric slicing strategies, i.e., \emph{2D-axial}, \emph{2D-axial+}, \emph{2.5D}, and \emph{2.5D+}, against our proposed \emph{omnidirectional} volume slicing method introduced in Section~\ref{sec:volume_slicing}. To ensure a fair comparison across all datasets and view configurations, our volume slicing strategy benchmark is solely based on identical MLP prediction heads comprising $100$k trainable parameters. The number of views is empirically set to $8$, $16$, and $24$, providing a reasonable tradeoff between computational efficiency and captured slice diversity. The corresponding results for any combination of MLP and the respective volume slicing strategy are summarized in Table~\ref{tab1}, and refer to all non-gray-colored rows. 

As visible from the results, the \emph{$2$D-axial} approach performs worse than all alternative slicing methods, likely because the majority of volumetric information is discarded when relying on a single slice. We nevertheless included this experiment as it establishes a meaningful lower bound for the performance evaluation of different volume slicing methods. As indicated by the results, incorporating multiple axial slices represented by the \emph{$2$D-axial+} approach improves performance by leveraging a greater amount of volumetric information.

Incorporating additional complementary perspectives, as in the \emph{2.5D} slicing strategy, leads to a clear improvement in performance. For example, the average AUROC computed across all datasets increases from 0.7335 for the \emph{2D-axial+} configuration (three slices) to 0.7681 for the \emph{2.5D} setup (three slices). This trend extends to the \emph{2.5D+} variant, which consistently outperforms its \emph{2D-axial+} counterpart for the same number of views. While the performance differences are more pronounced between $3$, $8$, and $16$ views, the gap narrows at $24$ views, suggesting that incorporating diverse perspectives is particularly beneficial when only a limited number of slices are available.

Beyond canonical plane slicing, our proposed \emph{omnidirectional} slicing strategy, which integrates both canonical and non-canonical orientations, further improves downstream performance. On average, across all evaluated configurations, the \emph{omnidirectional} approach consistently outperforms both the \emph{2D-axial+} and \emph{2.5D+} methods. Furthermore, for our volumetric slicing strategy, we observe a general trend across datasets. Increasing the number of views typically leads to higher performance. This relationship remains consistent for the \emph{omnidirectional} approach, with the sole exception of the kidney dataset, highlighting a clear positive correlation between the number of views and overall model performance. Furthermore, although the proposed \emph{omnidirectional} slicing strategy does not achieve the highest performance on every individual dataset, it consistently surpasses its \emph{2D-axial+} and \emph{2.5D+} counterparts for the same number of views when considering the mean performance across all datasets. This demonstrates the method’s robust and, in most cases, superior capability in capturing versatile volumetric information relevant for downstream classification.

\subsection{Impact of Graph Neural Network Prediction Head} \label{sec:tomographview_evaluation}

Following the benchmarking results in Section~\ref{subsec:volume_slicing_benchmarking}, where our \emph{omnidirectional} slicing strategy showed the strongest performance, we now turn to aggregating slice-level representations using the spherical mesh–based graph topology and its associated graph-based feature aggregation strategy introduced in Section~\ref{sec:graph_toplogy_construction}.
To evaluate the impact of the GNN-based prediction head, we compare two models: \mbox{\emph{MLP – Omnidirectional Slicing}} and \mbox{\emph{GNN – Omnidirectional Slicing}}, the latter constituting our full \emph{TomoGraphView} framework. Both methods operate on identical \emph{omnidirectional} slice inputs, ensuring that any performance differences arise solely from the aggregation method rather than the underlying representations. The corresponding results are summarized in Table~\ref{tab1}.


Our findings support the advantage of propagating information across slices via a meaningful underlying graph structure. Specifically, when comparing the average performance of \mbox{\emph{MLP - Omnidirectional Slicing}} against our \emph{TomoGraphView} framework (\emph{GNN - Omnidirectional Slicing}) across all six datasets, our proposed framework consistently outperforms the MLP for every view configuration, ultimately increasing from $0.8154$ to $0.8372$ in AUROC performance for the best performing model in each case. Notably, our \emph{TomoGraphView} framework achieves superior performance across nearly all dataset–view combinations, with the exception of Head-Neck at $8$ views and Soft-Tissue at $24$ views. These findings strongly highlight the effectiveness of the graph neural network-based feature aggregation approach over the MLP baseline method.

In summary, although our \emph{TomoGraphView} framework consistently improves the overall average performance across all experiments presented in Table~\ref{tab1}, it does not always achieve the single best score on every dataset, though it generally remains very close. We attribute these variations to reflect underlying dataset-specific characteristics, which we discuss further in Section~\ref{sec:discussion_graph-based-feature-aggregation}. 



\subsection{TomoGraphView - Impact of Volume Slice Spacing} \label{subsec:impact_volume_slice_spacing}

\begin{figure*}[t]
\centering
\includegraphics[width=1.0\linewidth]{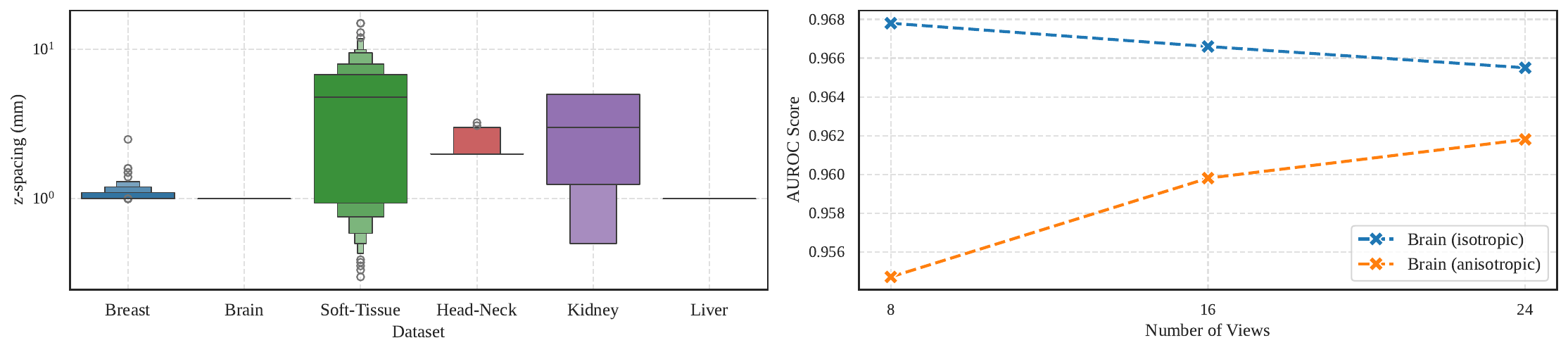}
\caption{Box plots illustrating the distribution of axial z-spacings across the oncology datasets used in this study (left), and the effect of slice spacing on the optimal number of views for TomoGraphView, shown for the brain tumors dataset using Area Under the Receiver Operating Characteristic Curve (AUROC) as the performance metric (right).}
\label{fig3:anisotropy}
\end{figure*}

A closer examination of the results obtained with the proposed \emph{TomoGraphView} framework, as shown in Table~\ref{tab1}, reveals distinct dataset-specific trends. For head-neck, kidney, liver, and soft-tissue tumors, the best performance is achieved when using the highest number of views. In contrast, for breast and brain tumor datasets, configurations with $8$ views yield superior results compared to those with a greater number of views. To better understand these patterns, we hypothesize that the observed performance differences with respect to the number of views are influenced by dataset-specific imaging characteristics, particularly the voxel spacing along the axial dimension. As illustrated in Figure~\ref{fig3:anisotropy}~(left), datasets with larger axial voxel spacing, such as soft-tissue, head-neck, and kidney, appear to benefit more from an increased number of views. This effect is reflected in the AUROC improvements observed when increasing the number of views from $8$ to $24$ in soft-tissue tumors (0.8399 vs. 0.8441), head and neck tumors (0.6780 vs. 0.7527), and kidney tumors (0.7878 vs. 0.7893).

To further validate this empirical observation, we conducted an additional experiment by artificially increasing the z-spacing in a high-resolution dataset. For this purpose, we utilize the brain tumor dataset, which features isotropic $1 \times 1 \times 1$ mm spacing and provides a sufficiently large sample size ($ n = 501$) for robust performance estimation. We resampled the isotropic samples to be anisotropic, specifically by increasing the axial spacing to $6$ mm, while keeping the other dimensions unchanged, and reevaluated our proposed \emph{TomoGraphView} framework for $8$, $16$, and $24$ views. 

As visualized in Figure~\ref{fig3:anisotropy} (right), we observe two effects: First, the overall performance decreased from the isotropic to the anisotropic setting, which is an expected consequence of information loss and resampling artifacts. Second, and more importantly, for the anisotropic setting, the performance across different view counts now follows the same trend as in high z-spacing datasets, such as soft-tissue, head-neck, or kidney, where increasing the number of views consistently improves results, as illustrated in Table~\ref {tab1}. This finding suggests that additional views are more suitable for compensating for the loss of information introduced by anisotropic resampling.

\subsection{TomoGraphView - Impact of Graph Topology} \label{sec:graph_topology_ablation}

\begin{figure}[t]
\centering
\includegraphics[width=1.0\linewidth]{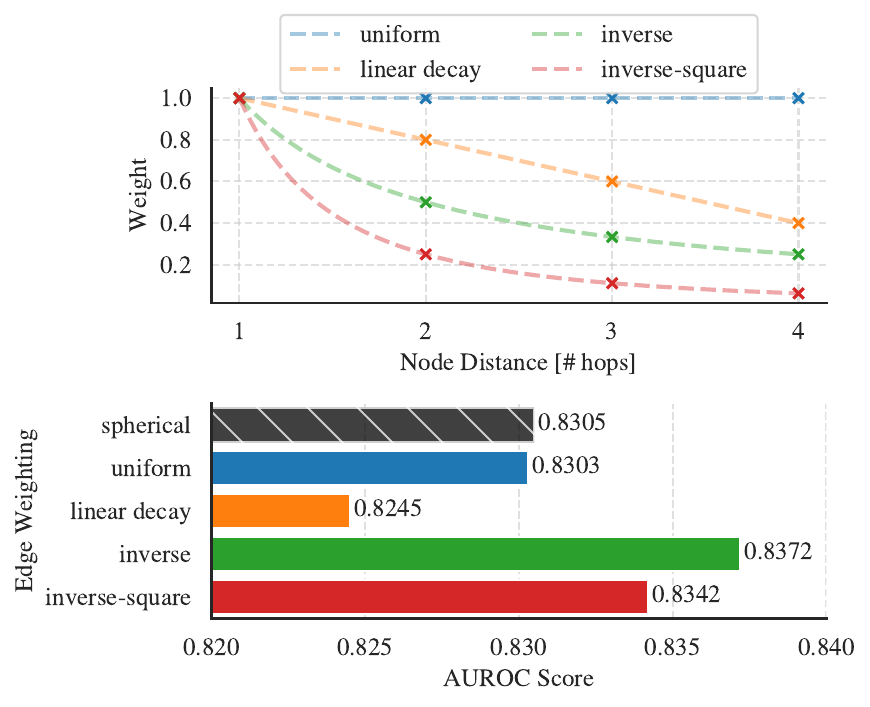}
\caption{Edge-weighting strategies based on node hop distance, including uniform, inverse, inverse-square, and linear-decay schemes (top). Corresponding graph topology performance for 24 views is shown using Area Under the Receiver Operating Characteristic Curve (AUROC) as the evaluation metric for each weighting scheme (bottom).}
\label{fig:graph_topology}
\end{figure}

We continue by studying the impact of the underlying graph topology, comparing different graph structures to assess the effect of graph connectivity on our proposed \emph{TomoGraphView} framework. In total, we compare five distinct graph topologies, which we refer to as \emph{spherical}, \emph{uniform}, \emph{linear decay}, \emph{inverse}, and \emph{inverse-square}. More specifically, \emph{uniform}, \emph{linear decay}, \emph{inverse}, and \emph{inverse-square} topologies share the same graph connectivity (i.e., complete graph), but show differences in the edge-weights for node cross-connections (i.e., connections to non-immediate neighbors after triangulation as indicated in Section~\ref{sec:graph_toplogy_construction}). The corresponding edge weights are determined according to the node distance in hops as indicated by Equation~\ref{eq:edge_weights_hop_distance} and depicted in Figure~\ref{fig:graph_topology}~(top). In contrast, the \emph{spherical} scheme does not use any node cross-connections, thus only connects nodes that are adjacent in the Delaunay triangulation (see Section~\ref{sec:graph_toplogy_construction}), effectively representing a localized graph structure. Hence, the \emph{spherical} topology results in a substantially sparser graph representation compared to the fully connected variants.

We conducted the graph topology experiments across all datasets and report the average AUROC performance using configurations with 24 nodes, as previous experiments in Section~\ref{sec:tomographview_evaluation} demonstrated this to be the overall best-performing setting for \emph{TomoGraphView}. Additionally, a larger number of nodes enables a more informative comparison, since the addition of edges has a greater relative effect in denser graphs. The corresponding results are presented in Figure~\ref{fig:graph_topology}~(bottom). 

Several key observations emerge from these experiments. First, the \emph{spherical} topology achieves comparable performance to the fully connected \emph{uniform} topology, despite relying on substantially fewer edge connections. This highlights the efficiency and relevance of the local spherical graph connectivity pattern. Second, introducing cross-connections between all nodes improves performance only when employing smooth distance-based weighting schemes such as \emph{inverse} and \emph{inverse-square}, which effectively expand the receptive field by aggregating information from diverse perspectives. Among these, the \emph{inverse} weighting outperforms the \emph{inverse-square} variant. In contrast, the \emph{linear-decay} weighting scheme, although also expanding the receptive field, leads to a decline in AUROC performance. This finding highlights the importance of using decaying weighting functions that place a stronger relative emphasis on local connections (i.e., one-hop neighbors) compared to global ones (i.e., node cross-connections). Detailed results, presented on a per-dataset basis for each of the different graph topologies, are provided in the~\ref{appendix:graph_topology_ablation}.

\subsection{Slice-wise Feature Aggregation Method Benchmark} \label{sec:slice-wise_feature_aggregation_methods}


In this section, we compare \emph{TomoGraphView} with two established slice-wise aggregation methods: a long short-term memory (LSTM) approach, which interprets a $3$D volume as a sequence of consecutive slice embeddings, and a transformer-based approach, represented by the medical slice transformer (MST)~\cite{muller2025medical}, which employs an attention mechanism to fuse intra-slice information. To ensure a fair comparison with previous experiments, the number of trainable parameters is again fixed to $100$k across all models. Additionally, we evaluate both LSTM- and transformer-based (MST) aggregation when combined with our proposed \emph{omnidirectional} slicing strategy, thereby disentangling the effects of the aggregation mechanism from those of slice selection.

Extending the MST-based aggregation to \emph{omnidirectional} slices does not require any explicit modeling of slice relationships, as the cross-attention mechanism inherently aims to capture inter-slice relationships. In contrast, this adaptation is less intuitive for LSTM models, which are inherently designed to process naturally sequential inputs. Since \emph{omnidirectional} slices, derived from points sampled on a spherical surface, lack a natural ordering, we impose and fix a consistent sequence based on node indices. The mean AUROC and MCC values comparing \emph{TomoGraphView}, LSTM, and MST across all datasets are summarized in Table~\ref{tab4}. 

We start by evaluating the LSTM and MST approaches in their original formulation using consecutive axial slices (i.e., 2D-axial+). The results show that the best-performing view setting ($16$) in the MST approach outperforms the best-performing view setting ($24$) in the LSTM approach, for both evaluation metrics, with AUROC values of $0.7768$ vs. $0.7446$ and MCC values of $0.3995$ vs. $0.3365$, highlighting the strong aggregation capability of the MST. Furthermore, this performance advantage is even observable across all view configurations, with the MST again surpassing the LSTM in both evaluation metrics. Nevertheless, both methods fall short of our proposed \emph{TomoGraphView} framework, which overall achieves an AUROC of $0.8372$ and an MCC of $0.5191$.

Admittedly, comparing LSTM and MST in their original 2D-axial+ slicing configuration to \emph{TomoGraphView} is only partially fair, as differences may not only arise from the aggregation mechanism but also from the contextual information available in the input slices. To address this, we further extend the evaluation by applying \emph{omnidirectional} slicing to both LSTM and MST. Two key observations emerge: (I) MST continues to outperform LSTM across all view configurations, and (II) both models experience substantial performance improvements. Specifically, for LSTM, \emph{omnidirectional} slicing increases AUROC from $0.7446$ to $0.7863$ and MCC from $0.3365$ to $0.4444$. Similarly, for MST, AUROC improves from $0.7768$ to $0.8198$ and MCC from $0.3967$ to $0.4732$. These results highlight the clear benefit and broad applicability of our proposed \emph{omnidirectional} volume slicing strategy. Remarkably, while both baselines show notable gains, \emph{TomoGraphView} still achieves the highest overall performance.

\begin{figure*}[t]
\centering
\includegraphics[width=1.0\linewidth]{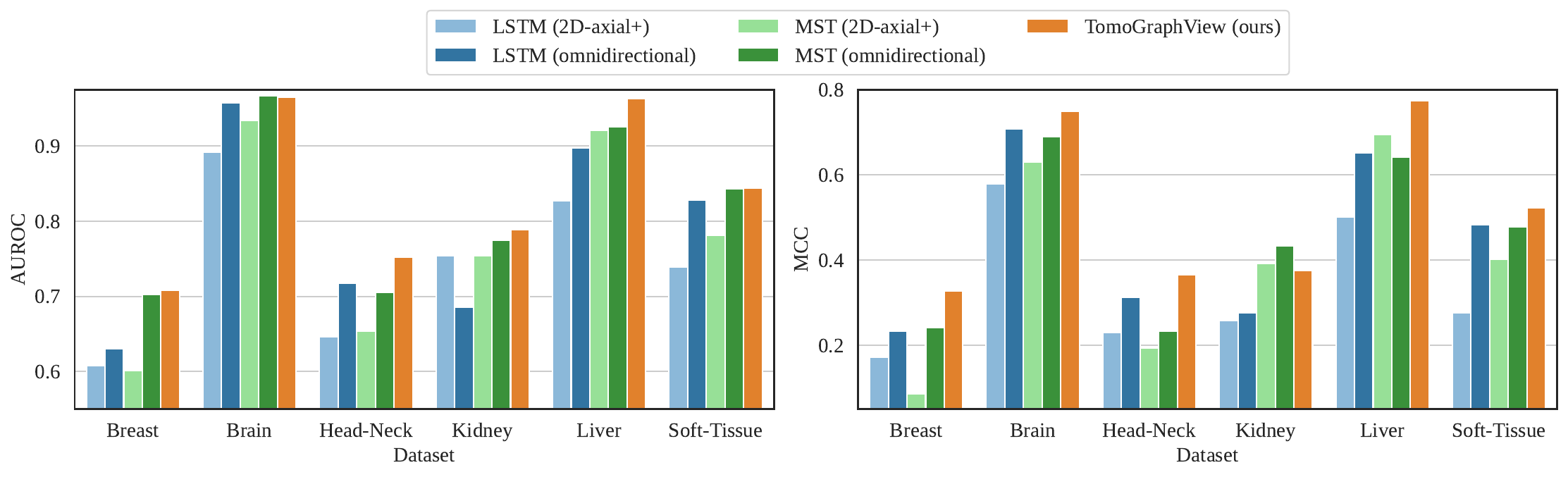}
\caption{Detailed results comparing slice-wise aggregation methods: long short-term memory (LSTM), medical slice transformer (MST), and TomoGraphView. Besides their original slicing
strategy (2D-axial+), LSTM, and MST are also evaluated using omnidirectional volume slicing. All methods are evaluated across all six datasets using the Area Under the Receiver Operating Characteristic Curve (AUROC) on the left and Matthews correlation coefficient (MCC) on the right as performance metrics. Bars indicate the average performance across five folds.}
\label{fig:slice_aggregation_results_per_dataset}
\end{figure*}

A more detailed, dataset-wise comparison is presented in Figure~\ref{fig:slice_aggregation_results_per_dataset}, which reports AUROC (left) and MCC (right) scores for $24$ views. Comprehensive numeric results across datasets, models, and view configurations are provided in Appendix Table~\ref{tab7}. Several trends can be observed: in terms of AUROC, \emph{TomoGraphView} matches the performance of the best baseline method on the breast, brain, and soft-tissue tumor datasets, but clearly surpasses them on the head-neck, kidney, and liver tumor datasets. A similar pattern holds for MCC, where \emph{TomoGraphView} consistently outperforms all competing methods, except for the kidney tumor dataset. Here, both MST variants show superior performance. Nevertheless, the overall results underscore the strong capability of \emph{TomoGraphView} for slice-wise feature aggregation, establishing it as both a competitive and compelling alternative to existing slice-wise approaches.



\begin{table}[t]
\centering

\caption{Quantitative results comparing a long short-term memory (LSTM) approach, medical slice transformer (MST), and our TomoGraphView framework for slice-wise feature aggregation performance. Besides their original slicing strategy (2D-axial+), LSTM and MST are also evaluated using \emph{omnidirectional} volume slicing. Methods are compared across different numbers of views and performance metrics, including the Area Under the Receiver Operating Characteristic Curve (AUROC) and the Matthews correlation coefficient (MCC). Results are presented as the average across all six datasets and five folds.}
\begin{adjustbox}{width=\linewidth}

\begin{tabular}{
    l |
    l |
    c |
    S[table-format=1.4]
    S[table-format=1.4]
}
\toprule
\textbf{Model} & \textbf{Volume Slicing} & \textbf{Views} & {\textbf{AUROC}} & {\textbf{MCC}} \\ 
\midrule
\multirow{3}{*}{LSTM} 
    & \multirow{3}{*}{2D-axial+} & 8  & 0.7305 & 0.3101 \\
    &                            & 16 & 0.7330 & 0.3129 \\
    &                            & 24 & 0.7446 & 0.3365 \\ 
\midrule
\multirow{3}{*}{LSTM} 
    & \multirow{3}{*}{Omnidirectional} & 8  & 0.7764 & 0.4225 \\
    &                                  & 16 & 0.7744 & 0.4178 \\
    &                                  & 24 & 0.7863 & 0.4444 \\ 
\midrule \midrule
\multirow{3}{*}{MST}  
    & \multirow{3}{*}{2D-axial+} & 8  & 0.7583 & 0.3442 \\
    &                            & 16 & 0.7768 & 0.3967 \\
    &                            & 24 & 0.7746 & 0.3995 \\ 
\midrule
\multirow{3}{*}{MST}  
    & \multirow{3}{*}{Omnidirectional} & 8  & 0.8083 & 0.4691 \\
    &                                  & 16 & 0.8171 & 0.4732 \\
    &                                  & 24 & 0.8198 & 0.4530 \\ 
\midrule \midrule
\multirow{3}{*}{TomoGraphView} 
    & \multirow{3}{*}{Omnidirectional} & 8  & \underline{0.8279} & 0.4891 \\
    &                                            & 16 & 0.8212 & \underline{0.5096} \\
    &                                            & 24 & \textbf{0.8372} & \textbf{0.5191} \\ 
\bottomrule
\end{tabular}
\end{adjustbox}
\label{tab4}
\end{table}

\subsection{Comparative Analysis with $3$D Pretrained Models} \label{sec:comparison_3D_models}

In the following, we compare our proposed TomoGraphView framework at $24$ views with large-scale pretrained 3D models. All baseline models are convolution-based, as 3D vision transformers are substantially more memory-demanding due to the much larger token sequences (i.e., number of patches) compared to 2D inputs. In addition, vision transformers typically require significantly larger datasets to pretrain effectively, whereas medical imaging datasets are often relatively limited in size.

The 3D models used in our comparison with TomoGraphView represent the top-performing downstream classification approaches identified by \citeauthor{aerts2025foundation}~\cite{aerts2025foundation} and \citeauthor{wald2024openmind}~\cite{wald2024openmind}. Specifically, we benchmark our framework against FMCIB~\cite{pai2024foundation}, Vista3D~\cite{he2025vista3d}, Models Genesis~\cite{zhou2019models}, as well as two pretrained models from \citeauthor{wald2024openmind}~\cite{wald2024openmind} based on VoCo~\cite{wu2024voco} and SwinUNETR~\cite{tang2022self}. A detailed description of these models is provided in~\ref{appendix:3d_pretrained_models}. 

The remainder of this section is structured as follows. In Section~\ref{subsec:frozen_backbones}, we first evaluate the fixed embeddings obtained from the selected large-scale pretrained 3D models across all datasets, without applying any task-specific finetuning to assess their representational capacity learned during pretraining. Subsequently, in Section~\ref{subsec:model_finetuning}, we perform dataset-specific finetuning, which typically improves downstream performance by allowing the pretrained representations to better adapt to the characteristics of each target task. Since all large-scale pretrained 3D models were trained in a self-supervised manner, we attach a classification head to the encoder to enable downstream classification. Consistent with our previous experiments, we fix the number of trainable parameters to $100$k to ensure a fair comparison. Notably, prediction heads previously discussed, such as GNN, transformer, or LSTM-based architectures, cannot be integrated with the $3$D pretrained models, as they require slice-based inputs rather than full 3D volume embeddings.

\subsubsection{TomoGraphView vs. Frozen 3D Backbones} \label{subsec:frozen_backbones} 

\begin{figure*}[t]
\centering
\includegraphics[width=1.0\linewidth]{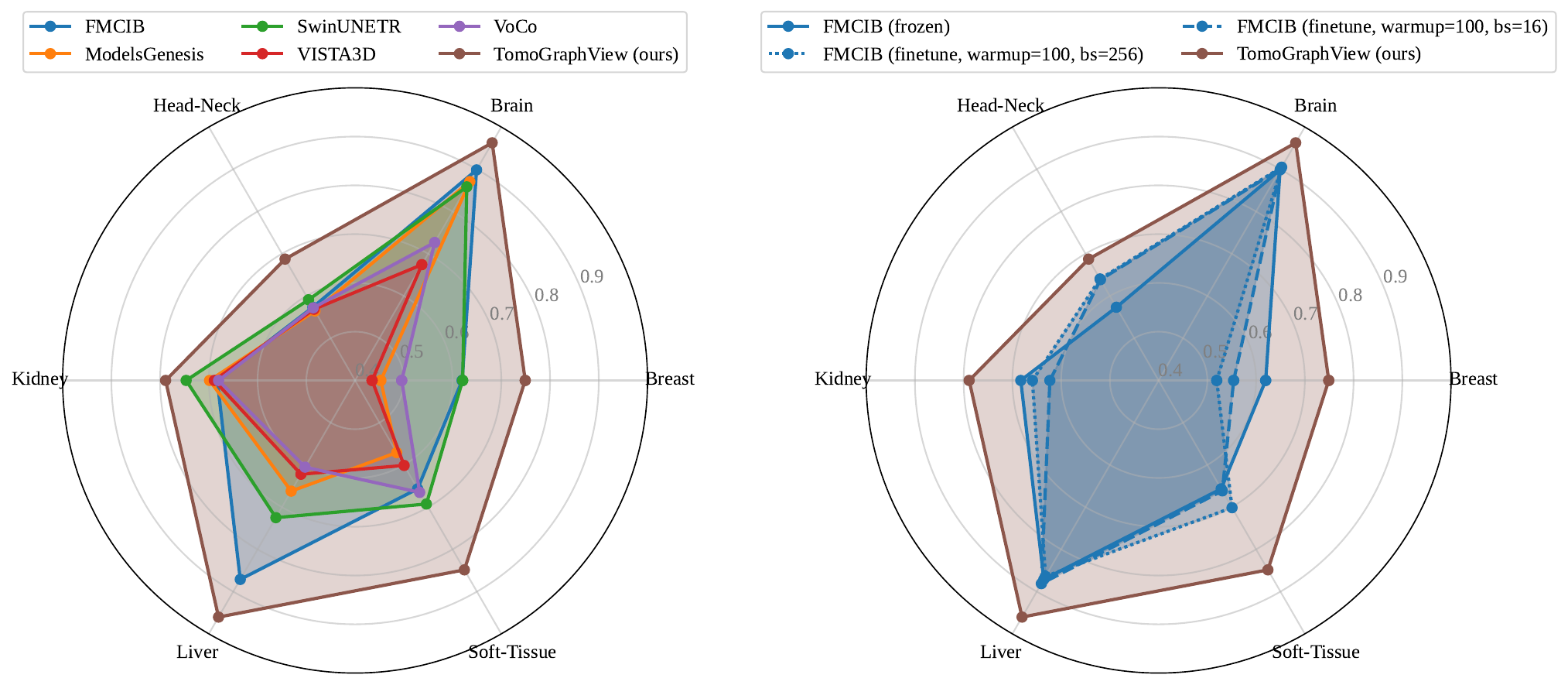}
\caption{Radar plots showing quantitative results comparing frozen $3$D backbones against our TomoGraphView framework (left) and fine-tuning results of the best performing $3$D backbone (FMCIB) against our TomoGraphView framework (right). Plots show the average Area Under the Receiver Operating Characteristic Curve (AUROC) performance across 5-folds.}
\label{fig:results_frozen_finetune_backbones}
\end{figure*}

We first evaluate the five large-scale pretrained 3D models across all six oncological downstream tasks and compare their performance with our proposed TomoGraphView framework. The results are summarized in the left radar plot of Figure~\ref{fig:results_frozen_finetune_backbones}. Comprehensive quantitative results for all methods, datasets, and performance metrics are provided in Appendix Table~\ref{tab5}.

Across all tasks, our proposed \emph{TomoGraphView} framework achieves an average AUROC of $0.8372$, outperforming the on average second-best approach, FMCIB~\cite{pai2024foundation}, which achieves an AUROC of $0.7170$. This demonstrates a consistent advantage of \emph{TomoGraphView} over all competing models. Looking more closely, on a per-task level, \emph{TomoGraphView} also yields the highest AUROC for each dataset: breast tumors ($0.7084$ vs. $0.6207$, SwinUNETR), brain tumors ($0.9655$ vs. $0.8985$, FMCIB), head–neck tumors ($0.7527$ vs. $0.5910$, SwinUNETR), kidney tumors ($0.7893$ vs. $0.7466$, SwinUNETR), liver tumors ($0.9634$ vs. $0.8711$, FMCIB), and soft-tissue tumors ($0.8441$ vs. $0.6926$, SwinUNETR). Although the performance gains are less pronounced for kidney tumors ($0.7893$ vs. $0.7466$) and brain tumors ($0.9655$ vs. $0.8985$), \emph{TomoGraphView} consistently outperforms all competing methods by a clear margin across the remaining breast, head-neck, liver, and soft-tissue tumor datasets. This observation, as exemplified by the AUROC results, can be further confirmed by additional evaluation metrics, such as balanced accuracy, F1-Score, and Matthews correlation coefficient, underscoring the great potential of \emph{TomoGraphView}. Detailed results can be found in the Appendix Table~\ref{tab5}.

\subsubsection{TomoGraphView vs. Finetuned 3D Backbones} \label{subsec:model_finetuning}

Building on the frozen backbone results presented in the previous subsection, where, on average, the FMCIB~\cite{pai2024foundation} approach emerged as the best-performing large-scale pretrained 3D model across all datasets and different performance metrics, we now evaluate its finetuning performance, as it allows the pretrained representations to better adapt to the specific characteristics of each downstream task. Strictly following the scheme introduced by \citeauthor{wald2024openmind}~\cite{wald2024openmind}, we finetune FMCIB for $200$ epochs in total, using a learning rate of $1\times10^{-4}$, a cosine annealing scheduler, a gradual warm-up of $100$ epochs applied to both the encoder and the classification head, and a batch size of $256$. Additionally, we explore a batch size of $16$, which is consistent with previous experiments. The corresponding results are illustrated in the right radar plot of Figure~\ref{fig:results_frozen_finetune_backbones}, with detailed quantitative results provided in Appendix Table~\ref{tab8}, which additionally shows decreased warm-up durations of $20$ and $50$ epochs to allow for more update steps under higher learning rates.

Overall, fine-tuning with a 100-epoch warm-up and a batch size of 256 yields performance gains across several downstream tasks. In particular, AUROC improves for brain tumors from $0.8985$ to $0.9049$, soft-tissue tumors from $0.6567$ to $0.7013$, and head–neck tumors from $0.5733$ to $0.6403$. However, no measurable improvements are observed for the breast, liver, or kidney tumor datasets compared to the frozen FMCIB backbone. Given that the datasets showing no improvements are also those with relatively small sample sizes, we conduct an additional experiment using a reduced batch size of $16$ to allow for more update steps during finetuning. This adjustment improves AUROC performance for the liver tumor dataset from $0.8711$ to $0.8811$, whereas performance for the breast and kidney datasets remains inferior. These results suggest that the effectiveness of fine-tuning is highly dataset-dependent, and a single universal configuration is unlikely to generalize across datasets without extensive hyperparameter exploration. Notably, our proposed \emph{TomoGraphView} framework continues to outperform the \mbox{FMCIB} model even under finetuning conditions.


        
\section{Discussion}



In this work, we introduce \emph{TomoGraphView}, a novel framework designed for $3$D medical image classification that preserves the spatial structure of volumetric medical imaging data while enabling compatibility with powerful $2$D vision foundation models. We comprehensively evaluate our approach across six diverse oncology datasets covering multiple anatomical regions (e.g., brain, breast, head-neck, kidney, or liver), various imaging modalities (CT and different MRI sequences), and various classification tasks primarily concerning tumor characterization. Building on our two key contributions, \emph{omnidirectional} volume slicing and \emph{spherical mesh-based feature aggregation}, our experimental evaluations show that \emph{TomoGraphView} achieves state-of-the-art results across downstream classification tasks, and even outperforms large-scale pretrained $3$D models. The following section provides a detailed discussion of the results and findings obtained throughout this work.

\subsection{Advancing Volume Slicing using Omnidirectional Views}

Generally, our experimental results shown in Table~\ref{tab1} suggest that more expressive tumor characterization, achieved by using richer and more representative volume slices from complementary directions, improves the description of the underlying 3D target structures. Specifically, expanding from purely axial $2$D slices to a $2.5$D multi-plane strategy consistently yields better downstream performance. As a result, the average AUROC performance across all evaluation datasets improves by $4.71\%$ (from $0.7335$ to $0.7681$) when comparing $2.5$D slicing with its corresponding $2$D-axial+ configuration (i.e., with three axial slices). This trend also persists for a larger number of views: with 24 views, AUROC increases by $2.39\%$ (from $0.7701$ to $0.7885$). 


However, as the first key contribution of this work, we presented a novel \emph{omnidirectional} volumetric slicing strategy that extends beyond the three conventional orientations axial, coronal, and sagittal and enables a richer characterization of volumetric structures that are not inherently aligned with these canonical axes. As a consequence, further gains are achieved with \emph{omnidirectional} volume slicing, which improves AUROC by $3.41\%$ (from $0.7885$ to $0.8154$) compared to $2.5D$+ slicing, and by $5.88\%$ (from $0.7701$ to $0.8154$) relative to pure $2$D-axial slicing. It is worth highlighting, that the benefit of \emph{omnidirectional} volume slicing generalizes beyond MLP prediction heads across various backbones, including LSTM-, transformer-, and GNN-based architectures, as outlined in Section~\ref{sec:slice-wise_feature_aggregation_methods}. Notably, although our \emph{omnidirectional} slicing strategy cannot be directly converted into a naturally meaningful slice ordering, as compared to consecutive axial slicing, LSTMs trained on \emph{omnidirectional} slices outperform those trained on consecutive axial slices, with a notable AUROC increase of $5.60\%$ (from $0.7446$ to $0.7863$). Similarly, transformer-based models show an AUROC improvement of $5.83\%$ (from $0.7746$ to $0.8198$).

While our omnidirectional volume slicing strategy consistently improves the average performance across all experiments in Table \ref{tab1}, it does not always yield the single best score for every individual dataset, although it generally remains very close. We attribute these differences as reflections of intrinsic dataset-specific characteristics. In particular, as all datasets offer their highest in-plane resolution when extracting axial slices (i.e., finer spacing in the coronal and sagittal directions), we examine the alignment of target structures along the z-axis. This axis exhibits the greatest anisotropy across all datasets and therefore governs the amount of low-resolution lesion information captured when slices are taken from predominantly coronal or sagittal orientations. To quantify anatomical z-axis alignment, we apply principal component analysis (PCA) to the 3D segmentation masks to identify each structure’s major axis. The resulting alignment score is computed as the absolute cosine similarity between this major axis and the anatomical z-axis, yielding a value between 0 and 1, corresponding to no alignment and perfect alignment, respectively. Z-axis alignment scores across all datasets are visible in the Appendix Table~\ref{tab9}. Overall, we observe the following trend: For datasets with high z-axis alignment such as head-neck (0.92) and soft-tissue tumors (0.93), \emph{omnidirectional} volume slicing strongly improves downstream performance, owing to its inherent slicing plane variability. However, datasets which show a lower z-axis alignment such as breast (0.40) and liver (0.10) tumors naturally benefit from several high-resolution axial slices as offered by e.g., $2.5$D+ volume slicing. 

Taken together, the strong results throughout this work show that omnidirectional slicing captures volumetric structures more effectively than conventional slicing strategies, leading to consistent, architecture-independent performance gains. These outcomes underscore the flexibility and added benefit of our approach as a compelling alternative to traditional methods such as $2$D-axial or $2.5$D slicing. We argue that omnidirectional slicing is highly relevant for the medical image analysis community working with slice-based pipelines, as it provides a simple yet elegant way to represent 3D targets in a spatially coherent manner. To facilitate adoption, we also provide a user-friendly Python library for omnidirectional volume slicing at \url{https://pypi.org/project/OmniSlicer}.


\subsection{Spatially Coherent Graph-Based Feature Aggregation} \label{sec:discussion_graph-based-feature-aggregation}

Beyond the choice of representing a volume as a set of $2$D slices, the way information from cross-sectional images is aggregated is of central importance. To this end, we presented our second core contribution of this work: GNN-based feature aggregation, which relies on a \emph{spherical mesh-based graph topology} to explicitly model the spatial relationships of encoded $2$D slices extracted from $3$D volumes. Our spherical graph design is derived from the \emph{omnidirectional} slicing strategy, enabling the integration of information across a set of in-plane and out-of-plane perspectives while explicitly encoding their spatial relationships within the graph topology. Our experimental evaluations in Section~\ref{sec:tomographview_evaluation} and Section~\ref{sec:slice-wise_feature_aggregation_methods} underscore the advantage of our graph-based feature aggregation method across datasets. 

While MLPs do not offer any built-in capability to model relationships among slices, LSTMs treat consecutive slices as temporal sequences, thereby preserving their spatial ordering within a volume. However, such an ordering is only naturally defined along a single plane of consecutive slices, restricting the representation to one direction and potentially overlooking complementary information from other orientations. Nevertheless, although \emph{omnidirectional} volume slicing improves LSTM downstream performance, there is no natural ordering of these slices, potentially hindering further gains. 

We attribute the superior performance of our GNN-based approach to its stronger inductive bias. Specifically, the node feature aggregation and node connectivity are predefined, whereas for transformers, such as MST-based aggregation, spatial relationships and global dependencies must be learned through their self-attention mechanism, introducing additional complexity. Moreover, transformers typically perform best in large-scale data settings, which are uncommon in medical imaging, limiting their practical application. We argue that simplifying the learning task by the topological constraints introduced in our GNN-based approach leads to improved overall average performance while handling limited data more efficiently.

Furthermore, we argue that the performance improvements also lie in the GNN’s ability to explicitly integrate both local and global information by propagating slice-level features across nodes through edge connections defined by the graph structure (see Section~\ref{sec:graph_toplogy_construction}). This mechanism is not available e.g., to the MLP, which instead processes all slice information through fully connected layers without explicit relational modeling. Furthermore, given that the employed graph topology approximates a three-dimensional geometric structure (i.e., a sphere), we argue that a graph-based representation is better suited to preserve the inherent spatial relationships among slices compared to dense connections in an MLP. Consequently, the graph structure provides a more faithful representation of the $3$D volume when represented through $2$D slices and their corresponding relative position.

\subsection{Balancing Data Dimensionality and Data Scarcity}

Current approaches to $3$D medical image classification face a fundamental trade-off between data dimensionality (2D slices vs. 3D volumes) and data quantity (i.e., sample size). Although $2$D slice-based methods achieve strong performance, they inherently lose anatomical context and spatial coherence when $3$D volumes are decomposed into individual slices. Conversely, models that directly process $3$D data preserve spatial structure but are constrained by the limited availability of pretraining data. Both, data dimensionality and data scarcity substantially impact a model’s representational capacity for volumetric analysis, two crucial aspects for reliable downstream classification. 

While foundation models have advanced rapidly in domains such as histopathology~\cite{filiot2024phikon, chen2024towards, koch2024dinobloom}, where large-scale $2$D data and sophisticated pretraining strategies drive strong transferability, radiology data-based $3$D foundation models remain comparatively underdeveloped~\cite{van2025foundation}. Our findings in Section~\ref{sec:comparison_3D_models} support this observation. 

We began by comparing \emph{TomoGraphView} with five large-scale pretrained $3$D models across six oncology datasets, using frozen encoder backbones. Evaluating raw embeddings enables a resource-efficient and faithful assessment of the representational quality learned during pretraining. Ideally, a robust pretrained model should produce embeddings that capture rich image semantics and correlate well with downstream performance. As \emph{TomoGraphView} also relies solely on frozen DINOv2 features, this comparison remains fair. Across all datasets, the FMCIB model~\cite{pai2024foundation} performs best among $3$D approaches but consistently trails \emph{TomoGraphView}, with relative improvements of 15.5\% in AUROC and 16.8\% in balanced accuracy (see Appendix Table~\ref{tab5}). 

Interestingly, despite FMCIB~\cite{pai2024foundation} being pretrained exclusively on body CT data (DeepLesion~\cite{yan2018deeplesion}), it even outperforms MRI-pretrained models on brain tumor MRI classification and matches the performance of the MRI-pretrained SwinUNETR~\cite{wald2024openmind}. We attribute this cross-modality generalization to the model’s ability to learn universal imaging features such as textures, edges, and spatial hierarchies. 

We further evaluated fine-tuning performance to assess how well FMCIB adapts pretrained representations to specific downstream tasks. Closely following the fine-tuning setup of \citeauthor{wald2024openmind}~\cite{wald2024openmind}, we observe improvements for brain, soft-tissue, and head–neck tumors, but no gain for breast, kidney, or liver datasets. This highlights that the effectiveness of fine-tuning is highly dataset-dependent and that no single configuration generalizes universally without extensive hyperparameter optimization. Nevertheless, overall fine-tuning performance still falls short of \emph{TomoGraphView}. In this study, we focused exclusively on finetuning large-scale pretrained $3$D models, as transfer learning has consistently demonstrated superior performance over training from scratch, particularly in medical image classification tasks where labeled data is often scarce. Existing studies show that leveraging prior knowledge from large-scale datasets enables faster convergence and improved generalization to domain-specific medical data~\cite{tajbakhsh2016convolutional}.

In summary, our results highlight the added value of combining $2$D foundation models with advanced volume-slicing and feature-aggregation strategies. We therefore position \emph{TomoGraphView} as a key component toward bridging the gap until fully native $3$D foundation models become available for medical image analysis.


\subsection{Limitations and Outlook}
While our proposed \emph{omnidirectional} volume slicing consistently improves performance across tasks, datasets, and prediction heads, it currently depends on the availability of tumor segmentation masks, as slice selection is based on identifying the largest lesion area. Notably, all other volumetric slicing methods in this study also rely on segmentation masks for this purpose. Fortunately, many established algorithms can generate sufficiently accurate tumor delineations, making our approach broadly applicable. Nevertheless, it is worth noting that $3$D medical image classification is usually performed on subvolumes, typically defined by bounding box coordinates around the structure of interest, to reduce the complexity of the problem. Given such a bounding box, we argue that selecting the central slice of the resulting subvolume should provide adequate information, thereby bypassing the need for precise tumor segmentation. Moreover, adding transparency, our \emph{omnidirectional} volume slicing strategy incurs a modest computational overhead during preprocessing, as generating $N$ \emph{omnidirectional} slices requires interpolating the volume $N$ times from different orientations. To address current limitations, future work will evaluate our framework on datasets providing bounding box annotations instead of segmentation masks. Furthermore, we plan to extend TomoGraphView by integrating a detection module, enabling a fully end-to-end 3D medical image classification pipeline. A promising direction in this regard is leveraging multimodal large language models, which can utilize in-context samples and textual prompts to predict bounding boxes for relevant structures, without requiring costly parameter updates or finetuning~\cite{ye2025multimodal}. Furthermore, as \emph{TomoGraphView} is designed as a flexible framework compatible with any $2$D foundation model, we anticipate that its performance will continue to improve in tandem with future advancements in $2$D backbone architectures and training paradigms. 

	
\section{Conclusion}

In this work, we present \emph{TomoGraphView}, a unified framework for $3$D medical image classification that combines \emph{omnidirectional} volume slicing with \emph{spherical graph-based feature aggregation}. Using six diverse datasets spanning multiple anatomical regions, imaging modalities, and clinical tasks, we demonstrated that (I) our \emph{omnidirectional} slicing approach consistently outperforms standard slicing strategies based on axial, sagittal, coronal, or combined views, and (II) \emph{spherical graph-based feature aggregation} improves performance beyond existing slice-wise feature aggregation baselines.

While $3$D models are naturally suited for tomographic data as they capture spatial context across all axes natively, their use in practice remains challenging: they require substantial computational resources, substantially larger datasets to achieve robust generalization due to their high parameter counts, and offer limited possibilities for transfer learning, as publicly available pretrained $3$D networks remain comparatively scarce and less established than $2$D alternatives \cite{van2025foundation}. As a result, slice-based $2$D methods remain an appealing alternative: they are more efficient to train, benefit from the large number of slices extracted from each volume, and can directly leverage powerful $2$D foundation models developed for natural images.

\emph{TomoGraphView} builds on these advantages. Our \emph{omnidirectional} slicing strategy provides a flexible and effective alternative to standard slicing along canonical planes, improving downstream performance across a wide range of prediction heads, including MLPs, LSTMs, and transformer-based approaches. Moreover, by integrating \emph{omnidirectional} volume slicing with spherical graph-based feature aggregation, our \mbox{\emph{TomoGraphView}} framework achieves superior performance levels when compared against both existing slice-wise feature aggregation methods and large-scale pretrained $3$D models. The results across our experimental evaluations underscore the value of combining powerful $2$D encoders with expressive methods capable of sampling and integrating versatile information from volumetric scans. Therefore, \emph{TomoGraphView} offers a practical and powerful approach as the field continues to progress toward fully native 3D foundation models.


In summary, this work highlights both the relevance and flexibility of $2$D slice-based strategies for $3$D medical image classification. By integrating powerful $2$D vision foundation models with improved volumetric slicing and graph-based aggregation, \emph{TomoGraphView} contributes a key step toward bridging the gap until robust and widely available $3$D foundation models become standard in medical image analysis.

\section*{Acknowledgments}

Authors JK, SMF, CIB, and JAS gratefully acknowledge financial funding received by Munich Center for Machine Learning, sponsored by the German Federal Ministry of Education and Research. Authors JCP and JK gratefully acknowledge funding from the Wilhelm Sander Foundation in Cancer Research (2022.032.1). Author SMF has received funding from the Deutsche Forschungsgemeinschaft (DFG, German Research Foundation) – 515279324 / SPP 2177.

\section*{Ethical Standards}

Ethical clearance for this study utilizing retrospectively collected in-house soft-tissue sarcoma data was obtained from the respective ethics committees of the data-providing institutions (i.e., Technical University of Munich, University of Washington/Seattle Cancer Care Alliance).

\section*{Data Availability}

All data, except the in-house soft-tissue sarcoma dataset, used in this study, are publicly available. We provide detailed descriptions and proper citations for each dataset used to ensure full transparency and reproducibility. Additionally, all code for data preprocessing and preparation is publicly released and can be used to fully reproduce our results. The code is available at \url{http://github.com/compai-lab/2025-MedIA-kiechle}. The in-house soft-tissue sarcoma data may be obtained from the corresponding author upon reasonable request and may be published online in the future after approval by the institutional ethics review boards. Furthermore, we provide a user-friendly library for \emph{omnidirectional} volume slicing at \url{https://pypi.org/project/OmniSlicer}.

\section*{Conflicts of Interest}
JCP holds shares in Mevidence and has received honoraria from AstraZeneca and Brainlab, all of which are outside the scope of the submitted work. The remaining authors declare that they have no known conflict of interest nor competing financial interests or personal relationships that could have appeared to influence the work reported in this paper.

\section*{Declaration of AI Assistance}
The authors acknowledge the use of ChatGPT to assist in improving the readability and language of this manuscript. The authors thoroughly reviewed and edited the text following its use and accept full responsibility for the content and integrity of the final publication.

\section*{Authorship Contribution Statement}
\textbf{Johannes Kiechle:} Writing – review \& editing, Writing – original draft, Visualization, Validation, Software, Resources, Project administration, Methodology, Investigation, Formal analysis, Data curation, Conceptualization. \textbf{Stefan M. Fischer:} Validation, Writing - Review \& Editing. \textbf{Daniel M. Lang:} Validation, Writing - Review \& Editing. \textbf{Cosmin I. Bercea:} Validation, Writing - Review \& Editing. \textbf{ Matthew J. Nyflot:} Data, Validation, Writing - Review \& Editing. \textbf{Lina Felsner:} Validation, Writing - Review \& Editing. \textbf{Julia A. Schnabel:} Validation, Writing - Review \& Editing, Supervision. \textbf{Jan C. Peeken:} Validation, Writing - Review \& Editing, Supervision, Funding acquisition


\bibliographystyle{elsarticle-harv}
\bibliography{bibliography}

\clearpage

\appendix

\section{Volume Slicing Comparison} \label{appendix:volume_slice_comparison_24_views}

While Figure~\ref{fig:volume_slicing} in Section~\ref{sec:volume_slicing} shows the resulting eight cross-sectional slices for a synthetically generated input volume, which displays a target structure that is not naturally aligned with canonical axes, we extend the visualization with 24 views in Figure~\ref{fig:volume_slicing_appendix} below. While the slice variability for both, 2D-axial+ (top) and 2.5D+ (middle), is limited, our \emph{omnidirectional} volume slicing strategy (bottom) excels by representing the target structure with all its facets, thereby depicting a more faithful representation of the target 3D structure.

\begin{figure*}[t]
\centering
\includegraphics[width=1.0\linewidth]{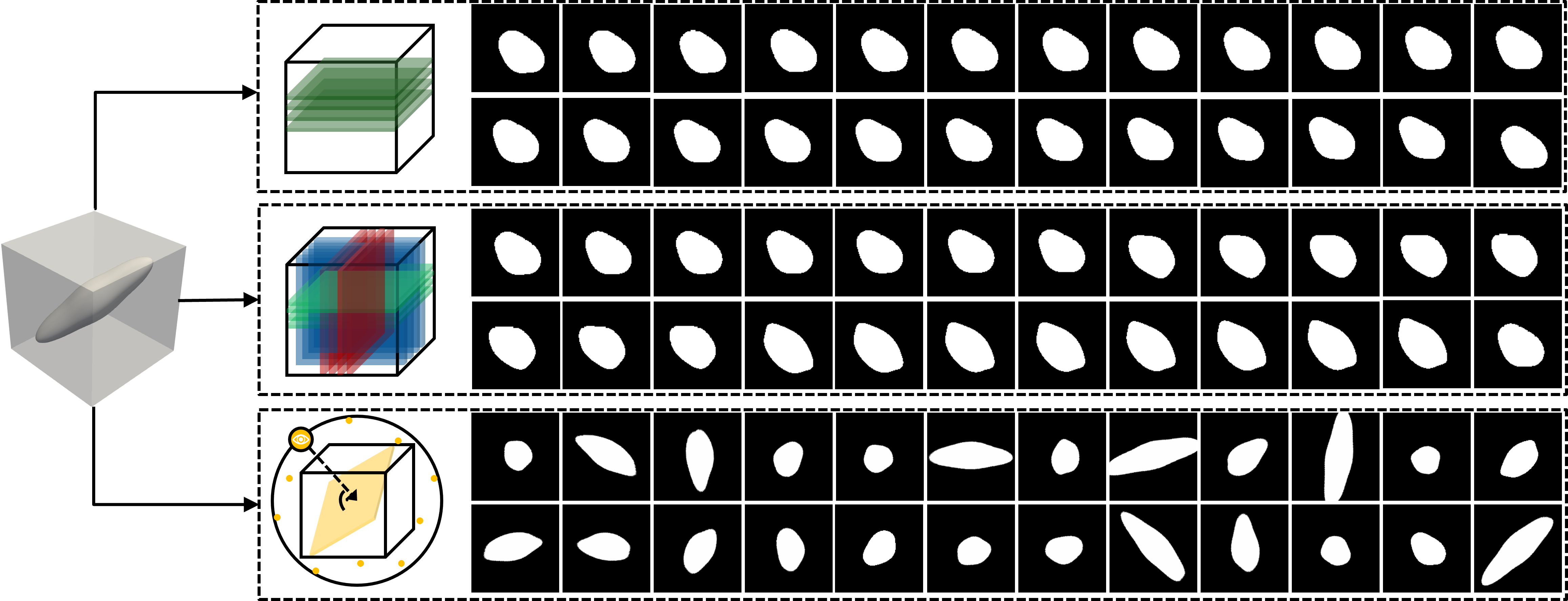}
\caption{A visual comparison of the three different volume slicing techniques and their resulting cross-sectional slices for a synthetically generated input volume, showing a target structure of interest which is not aligned with any canonical axis (i.e, axial, coronal, sagittal). The plot shows 2D-axial+ slicing (top), 2.5D+ slicing (middle), and our proposed \emph{omnidirectional} slicing (bottom), for 24 views each.}
\label{fig:volume_slicing_appendix}
\end{figure*}

\section{Dataset Task Descriptions} \label{appendix:dataset_task_description}

For all datasets except the head-neck and liver cohorts, this study focuses on tumor grading, where cases are categorized into entities of differing malignancy. Generally, tumor grade provides critical insight into how aggressive a tumor is and its potential for progression or metastasis, thereby serving as an essential factor for cancer staging and treatment decisions. For the liver dataset, we perform tumor subtype prediction, whereas for the Head-Neck dataset, we instead predict human papillomavirus (HPV) status, as HPV infection constitutes a major risk factor for oropharyngeal cancer (OPC). Importantly, HPV-positive OPCs have been shown to exhibit greater radio-sensitivity and lower cancer-related mortality compared to HPV-negative cases. Thus, determining HPV status constitutes a key diagnostic marker with direct implications for prognosis and treatment decisions. 

\section{$3$D Pretrained Model Description} \label{appendix:3d_pretrained_models}

\textbf{Models Genesis}~\cite{zhou2019models} is a collection of models built from unlabeled $3$D imaging data using a restorative reconstruction-based self-supervised method. It aims to generate powerful application-specific target models through transfer learning. The models demonstrate strong performance across various medical imaging tasks, emphasizing their potential for broad applicability in clinical settings.

\vspace{0.1cm}

\textbf{VOCO}~\cite{wu2024voco} is a large-scale $3$D medical image pre-training framework that leverages geometric context priors to learn consistent semantic representations. It is built on a substantial dataset of CT volumes and employs a novel pretext task for contextual position predictions. VOCO has demonstrated superior performance across various downstream tasks, establishing itself as a leading model in the field of medical imaging.

\vspace{0.1cm}

\textbf{VISTA$3$D}~\cite{he2025vista3d} is a versatile imaging segmentation and annotation model that supports both automatic and interactive segmentation of $3$D medical images. It is the first unified foundation model to achieve state-of-the-art performance across 127 classes and is designed to facilitate efficient human correction through its interactive features. VISTA$3$D integrates a novel supervoxel method to enhance zero-shot performance, marking a significant advancement in $3$D medical imaging.

\vspace{0.1cm}

\textbf{FMCIB}~\cite{pai2024foundation} is a foundation model designed to distinguish between lesions and non-lesions at the patch level in medical imaging. It aims to enhance the detection and characterization of cancerous lesions by leveraging self-supervised learning techniques. It is based on SimCLR~\cite{chen2020simclr} (Simple Framework for Contrastive Learning of Visual Representations), a self-supervised learning approach that utilizes contrastive learning to pretrain deep neural networks without the need for labeled data. The core idea behind SimCLR is to maximize the similarity between differently augmented views of the same image while minimizing the similarity between views of different images.

\vspace{0.1cm}

\textbf{SwinUNETR}~\cite{cao2022swin} was proposed as a pretraining method for the identically named SwinUNETR architecture and is composed of three components: image inpainting, Rotation prediction, and a contrastive training objective. The inpainting itself is a simple L$1$ loss applied to a masked-out image region, the rotation is a rotation of 0°, 90°, 180°, or 270° degrees along the z-axis, with an MLP used to classify the applied rotation. Lastly, the contrastive coding ensures that the linearly projected representations of the encoder are highly similar or dissimilar, depending on whether the two sub-volumes belong to the same or a different image, respectively. These three losses are combined with an equal weighting to form the SwinUNETR pretraining.

\section{Graph Topology Ablation} \label{appendix:graph_topology_ablation}

Detailed quantitative results of our graph topology ablation experiment performed in Section~\ref{sec:graph_topology_ablation} can be found in Table~\ref{tab6}. In this context, we compare our proposed \emph{TomoGraphView} framework across various graph topologies and edge connection weighting schemes. For graph topology, we differentiate between local and complete, where local represents a pure spherical structure, while complete represents the spherical structure with additional cross-connections between all nodes in the graph. To this end, we evaluate different edge weighting schemes, including uniform, linear decay, inverse, and inverse-square, which indicate decaying cross-node weights with increasing hop distances. 

\begin{table*}[t]
\centering
\caption{Quantitative results for the graph topology analysis comparing different graph topologies and edge weighting schemes. Results are represented as the mean Area under the Receiver Operating Characteristic Curve (AUROC) value across 5 folds. The best-performing method is indicated in bold, while the second-best is underlined. The column 'Mean' averages the performance across all datasets.}

\begin{tabular}{
    l |
    c |
    c |
    S[table-format=1.4]
    S[table-format=1.4]
    S[table-format=1.4]
    S[table-format=1.4]
    S[table-format=1.4]
    S[table-format=1.4] |
    S[table-format=1.4]
}
\toprule
\textbf{Method} & \textbf{Topology} & \textbf{Weighting} &
\textbf{Breast} & \textbf{Brain} & \textbf{Head–Neck} &
\textbf{Kidney} & \textbf{Liver} & \textbf{Soft-Tissue} & \textbf{Mean} \\
\midrule

\multirow{5}{*}{TomoGraphView}
 & local    & uniform        & 0.6997 & \underline{0.9667} & \textbf{0.7770} & 0.7526 & \textbf{0.9666} & 0.8204 & 0.8305 \\
 & complete & uniform        & 0.6939 & \textbf{0.9668} & 0.7480 & 0.7672 & 0.9582 & \textbf{0.8475} & 0.8303 \\
 & complete & linear decay   & 0.6857 & 0.9653 & 0.7503 & 0.7474 & 0.9614 & 0.8371 & 0.8245 \\
 & complete & inverse        & \underline{0.7084} & 0.9655 & 0.7527 & \textbf{0.7893} & 0.9634 & \underline{0.8441} & \textbf{0.8372} \\
 & complete & inverse-square & \textbf{0.7111} & 0.9621 & \underline{0.7549} & \underline{0.7878} & \underline{0.9646} & 0.8249 & \underline{0.8342} \\

\bottomrule
\end{tabular}
\label{tab6}
\end{table*}

\section{Slice-wise Feature Aggregation Methods} \label{appendix:slice-wise_feature_agggregation}

Detailed quantitative results of our slice-wise feature aggregation benchmark experiment performed in Section~\ref{sec:slice-wise_feature_aggregation_methods} can be found in Table~\ref{tab7}. Therein, we compare different slice-wise feature aggregation methods, including long short-term memory (LSTM) and medical slice transformer (MST), with our \emph{TomoGraphView} framework across various volume slicing strategies and numbers of views. Numbers are indicated as the mean Area under the Receiver Operating Characteristic Curve (AUROC) value across 5 folds. 

\begin{table*}[t]
\centering
\caption{Quantitative results for the slice-wise feature aggregation benchmark comparing different methods, such as long short-term memory (LSTM) and medical slice transformer (MST) with \emph{TomoGraphView} across volume slicing strategies (i.e., 2D-axial+ and \emph{omnidirectional}) and number of views (i.e., 8, 16, and 24). Results are represented as the mean Area under the Receiver Operating Characteristic Curve (AUROC) value across 5 folds. The best-performing method is indicated in bold, while the second-best is underlined. The column 'Mean' averages the performance across all datasets.}

\begin{tabular}{
    l |
    l |
    c |
    S[table-format=1.4]
    S[table-format=1.4]
    S[table-format=1.4]
    S[table-format=1.4]
    S[table-format=1.4]
    S[table-format=1.4] |
    S[table-format=1.4]
}
\toprule
\textbf{Method} & \textbf{Volume Slicing} & \textbf{Views} &
\textbf{Breast} & \textbf{Brain} & \textbf{Head–Neck} &
\textbf{Kidney} & \textbf{Liver} & \textbf{Soft-Tissue} & \textbf{Mean} \\
\midrule

\multirow{3}{*}{LSTM} & \multirow{3}{*}{2D-axial+} 
 & 8  & 0.6437 & 0.8826 & 0.6048 & 0.6669 & 0.8923 & 0.7254 & 0.7305 \\
 & & 16 & 0.6505 & 0.8957 & 0.6207 & 0.7072 & 0.7866 & 0.7374 & 0.7330 \\
 & & 24 & 0.6077 & 0.8922 & 0.6467 & 0.7546 & 0.8275 & 0.7388 & 0.7446 \\
\midrule

\multirow{3}{*}{LSTM} & \multirow{3}{*}{Omnidirectional} 
 & 8  & 0.6980 & 0.9419 & 0.6510 & 0.6705 & 0.9133 & 0.7837 & 0.7764 \\
 & & 16 & 0.6731 & 0.9497 & 0.6670 & 0.6733 & 0.9137 & 0.7696 & 0.7744 \\
 & & 24 & 0.6303 & 0.9580 & 0.7179 & 0.6861 & 0.8976 & 0.8280 & 0.7863 \\
\midrule\midrule

\multirow{3}{*}{MST} & \multirow{3}{*}{2D-axial+} 
 & 8  & 0.6497 & 0.9232 & 0.5949 & 0.7124 & 0.9002 & 0.7696 & 0.7583 \\
 & & 16 & 0.6539 & 0.9308 & 0.6294 & 0.7563 & 0.9081 & 0.7824 & 0.7768 \\
 & & 24 & 0.6014 & 0.9346 & 0.6539 & 0.7558 & 0.9206 & 0.7815 & 0.7746 \\
\midrule

\multirow{3}{*}{MST} & \multirow{3}{*}{Omnidirectional} 
 & 8  & 0.7037 & 0.9655 & 0.6443 & 0.7573 & 0.9497 & 0.8295 & 0.8083 \\
 & & 16 & \underline{0.7164} & 0.9663 & 0.7225 & 0.7286 & 0.9238 & \textbf{0.8452} & 0.8171 \\
 & & 24 & 0.7029 & \underline{0.9669} & 0.7050 & 0.7751 & 0.9254 & 0.8434 & 0.8198 \\
\midrule\midrule

\multirow{3}{*}{TomoGraphView} & \multirow{3}{*}{Omnidirectional} 
 & 8  & \textbf{0.7336} & \textbf{0.9678} & 0.6780 & \underline{0.7878} & \underline{0.9600} & 0.8399 & 0.8279 \\
 & & 16 & 0.6907 & 0.9666 & \underline{0.7244} & 0.7479 & 0.9585 & 0.8389 & \underline{0.8212} \\
 & & 24 & 0.7084 & 0.9655 & \textbf{0.7527} & \textbf{0.7893} & \textbf{0.9634} & \underline{0.8441} & \textbf{0.8372} \\
\bottomrule
\end{tabular}
\label{tab7}
\end{table*}

\section{$3$D Pretrained Backbones Results} \label{appendix:3d_backbones_results}

Detailed quantitative results of our comparative large-scale $3$D pretrained model analysis performed in Section~\ref{sec:comparison_3D_models} can be found in Table~\ref{tab5} (frozen backbone performance) and \mbox{Table~\ref{tab8}} (finetuning performance). 

\begin{table*}[t]
\centering
\caption{Quantitative results for the frozen backbone benchmark comparing different methods relying on different pretraining schemes, such as Foundation Model for Cancer Imaging Biomarkers (FMCIB), Models Genesis, SwinUNETR, Versatile Imaging SegmenTation and Annotation model (VISTA3D), Volume Contrast (VoCo) with \emph{TomoGraphView}. Results are represented as the mean Area under the Receiver Operating Characteristic Curve (AUROC), balanced accuracy (ACC), F1-score, and Matthews correlation coefficient (MCC), across 5 folds. The best-performing method per dataset is indicated in bold, while the second-best is underlined. The last section, indicated as 'Mean', averages the performance across all datasets and performance metrics.}

\begin{tabularx}{\textwidth}{
    X |
    c |
    X |
    c
    c
    c
    c
}
\toprule
\textbf{Dataset} & \textbf{Modality} & \textbf{Method} & {\textbf{AUROC}} & {\textbf{ACC}} & {\textbf{F1-Score}} & {\textbf{MCC}} \\
\midrule

\multirow{6}{*}{\textbf{Breast Tumors}} 
 & \multirow{6}{*}{MRI} & FMCIB & 0.6197 & 0.5427 & 0.5057 & \underline{0.1016} \\
 &  & ModelsGenesis & 0.4527 & 0.5017 & 0.3537 & 0.0024 \\
 &  & SwinUNETR & \underline{0.6207} & \underline{0.5445} & \underline{0.5303} & 0.0876 \\
 &  & VISTA3D & 0.4349 & 0.4592 & 0.3950 & -0.0844 \\
 &  & VoCo & 0.4957 & 0.5149 & 0.5049 & 0.0321 \\
 &  & TomoGraphView (ours) & \textbf{0.7084} & \textbf{0.6573} & \textbf{0.6553} & \textbf{0.3177} \\

\midrule
\multirow{6}{*}{\textbf{Brain Tumors}} 
 & \multirow{6}{*}{MRI} & FMCIB & \underline{0.8985} & \underline{0.7921} & \underline{0.8587} & \underline{0.5716} \\
 &  & ModelsGenesis & 0.8708 & 0.6257 & 0.7895 & 0.3037 \\
 &  & SwinUNETR & 0.8583 & 0.7694 & 0.7979 & 0.4646 \\
 &  & VISTA3D & 0.6741 & 0.5505 & 0.5843 & 0.0795 \\
 &  & VoCo & 0.7267 & 0.6719 & 0.7962 & 0.3569 \\
 &  & TomoGraphView (ours) & \textbf{0.9655} & \textbf{0.9131} & \textbf{0.9066} & \textbf{0.7443} \\

\midrule
\multirow{6}{*}{\textbf{Head–Neck Tumors}} 
 & \multirow{6}{*}{CT} & FMCIB & 0.5733 & 0.5407 & 0.5448 & 0.1027 \\
 &  & ModelsGenesis & 0.5645 & 0.5040 & 0.2805 & 0.0240 \\
 &  & SwinUNETR & \underline{0.5910} & 0.5609 & 0.5532 & 0.1424 \\
 &  & VISTA3D & 0.5686 & 0.5258 & 0.4875 & 0.0452 \\
 &  & VoCo & 0.5719 & \underline{0.5759} & \underline{0.6035} & \underline{0.1499} \\
 &  & TomoGraphView (ours) & \textbf{0.7527} & \textbf{0.6464} & \textbf{0.6711} & \textbf{0.2924} \\

\midrule
\multirow{6}{*}{\textbf{Kidney Tumors}} 
 & \multirow{6}{*}{CT} & FMCIB & 0.6828 & 0.6392 & 0.6785 & 0.2706 \\
 &  & ModelsGenesis & 0.6985 & 0.5667 & 0.5181 & 0.1279 \\
 &  & SwinUNETR & \underline{0.7466} & \underline{0.7112} & \textbf{0.7520} & \underline{0.4282} \\
 &  & VISTA3D & 0.6888 & 0.6102 & 0.5508 & 0.2614 \\
 &  & VoCo & 0.6801 & 0.6535 & 0.6973 & 0.3111 \\
 &  & TomoGraphView (ours) & \textbf{0.7893} & \textbf{0.7245} & \underline{0.7291} & \textbf{0.4412} \\

\midrule
\multirow{6}{*}{\textbf{Liver Tumors}} 
 & \multirow{6}{*}{CT} & FMCIB & \underline{0.8711} & \underline{0.8082} & \underline{0.8060} & \underline{0.6208} \\
 &  & ModelsGenesis & 0.6621 & 0.5695 & 0.5013 & 0.1743 \\
 &  & SwinUNETR & 0.7249 & 0.6696 & 0.6641 & 0.3406 \\
 &  & VISTA3D & 0.6221 & 0.5962 & 0.5941 & 0.2040 \\
 &  & VoCo & 0.6053 & 0.5896 & 0.5812 & 0.1802 \\
 &  & TomoGraphView (ours) & \textbf{0.9634} & \textbf{0.8956} & \textbf{0.8959} & \textbf{0.7939} \\

\midrule
\multirow{6}{*}{\textbf{Soft-Tissue Tumors}} 
 & \multirow{6}{*}{MRI} & FMCIB & 0.6567 & 0.5882 & 0.6499 & 0.2010 \\
 &  & ModelsGenesis & 0.5708 & 0.5432 & 0.5308 & 0.0928 \\
 &  & SwinUNETR & \underline{0.6926} & 0.5595 & 0.6841 & 0.1583 \\
 &  & VISTA3D & 0.6013 & 0.5380 & 0.6495 & 0.0897 \\
 &  & VoCo & 0.6651 & \underline{0.6327} & \underline{0.7151} & \underline{0.2737} \\
 &  & TomoGraphView (ours) & \textbf{0.8441} & \textbf{0.7340} & \textbf{0.7957} & \textbf{0.4815} \\

\midrule
\midrule
\multirow{6}{*}{\textbf{Mean}} 
 & \multirow{6}{*}{---} & FMCIB & \underline{0.7170} & \underline{0.6518} & \underline{0.6739} & \underline{0.3114} \\
 &  & ModelsGenesis & 0.6366 & 0.5518 & 0.4956 & 0.1209 \\
 &  & SwinUNETR & 0.7057 & 0.6358 & 0.6636 & 0.2703 \\
 &  & VISTA3D & 0.5983 & 0.5466 & 0.5435 & 0.0992 \\
 &  & VoCo & 0.6241 & 0.6064 & 0.6497 & 0.2173 \\
 &  & TomoGraphView (ours) & \textbf{0.8372} & \textbf{0.7618} & \textbf{0.7756} & \textbf{0.5191} \\
\bottomrule
\end{tabularx}
\label{tab5}
\end{table*}

\begin{table*}[t]
\centering
\small
\caption{Quantitative results for the finetuning performance of the best-performing frozen backbone, Foundation Model for Cancer Imaging Biomarkers (FMCIB). We employ different warmup durations, along with various batch sizes (BS), and compare their performance with our \emph{TomoGraphView} framework. Results are represented as the mean Area under the Receiver Operating Characteristic Curve (AUROC) value across 5 folds. The best-performing method is indicated in bold, while the second-best is underlined. The column 'Mean' averages the performance across all datasets.}

\begin{tabular}{
    l |
    c |
    c |
    c |
    S[table-format=1.4]
    S[table-format=1.4]
    S[table-format=1.4]
    S[table-format=1.4]
    S[table-format=1.4]
    S[table-format=1.4] |
    S[table-format=1.4]
}
\toprule
\textbf{Model} & \textbf{Warmup} & \textbf{BS} & \textbf{Params} & \textbf{Brain} & \textbf{Soft-Tissue} & \textbf{Breast} & \textbf{Head–Neck} & \textbf{Kidney} & \textbf{Liver} & \textbf{Mean} \\
\midrule

FMCIB (frozen) & n/a & 16 & 100k & 0.8985 & 0.6567 & \underline{0.6197} & 0.5733 & \underline{0.6828} & 0.8711 & \underline{0.7170} \\

\midrule
\multirow{4}{*}{FMCIB (finetune)} 
 & 20  & 256 & 184M & \underline{0.9147} & \underline{0.7322} & 0.5049 & 0.6286 & 0.5730 & 0.8708 & 0.7040 \\
 & 50  & 256 & 184M & 0.8981 & 0.7230 & 0.5050 & 0.6308 & 0.6162 & 0.8636 & 0.7061 \\
 & 100 & 256 & 184M & 0.9049 & 0.7013 & 0.5193 & \underline{0.6403} & 0.6589 & 0.8639 & 0.7147 \\
 & 100 & 16  & 184M & 0.9016 & 0.6615 & 0.5540 & 0.6386 & 0.6231 & \underline{0.8811} & 0.7100 \\

\midrule
TomoGraphView (ours) & n/a & 16 & 100k & \textbf{0.9626} & \textbf{0.8485} & \textbf{0.7491} & \textbf{0.6873} & \textbf{0.7887} & \textbf{0.9603} & \textbf{0.8328} \\

\bottomrule
\end{tabular}
\label{tab8}
\end{table*}

\begin{table*}[t]
\centering
\caption{Quantitative dataset-specific metrics, showing median z-axis spacing, its corresponding z-axis spacing interquartile range (IQR) and z-axis alignment of the target anatomical structure. To quantify anatomical z-axis alignment, we apply principal component analysis (PCA) to the 3D segmentation masks to identify each structure’s major axis. The resulting alignment score is computed as the absolute cosine similarity between this major axis and the anatomical z-axis, yielding a value between 0 (no alignment) and 1 (perfect alignment). The unit measuring median slice spacing and IQR slice spacing is given in millimeters (mm).}
\begin{tabularx}{\textwidth}{X|c|cccccc}
\toprule
\textbf{Metric} 
& \textbf{Unit} 
& \textbf{Breast} 
& \textbf{Brain} 
& \textbf{Head-Neck} 
& \textbf{Kidney} 
& \textbf{Liver} 
& \textbf{Soft-Tissue} \\
\midrule
z-axis spacing (median)    & mm & 1.00 & 1.00 & 2.00 & 3.00 & 1.00 & 4.80 \\
z-axis spacing (IQR)       & mm & 0.10 & 0.00 & 0.00 & 3.75 & 0.00 & 5.86 \\
z-axis alignment       & n/a  & 0.40 & 0.32 & 0.92 & 0.51 & 0.10 & 0.93 \\ \midrule
\bottomrule
\end{tabularx}

\label{tab9}
\end{table*}

\end{document}